\documentclass[aps,pra,twocolumn,amsmath,amssymb,groupedaddress,longbibliography]{revtex4-2}
\usepackage{graphicx}
\usepackage{dcolumn}
\usepackage{bm}
\usepackage[pdfstartview=FitH, CJKbookmarks=true, bookmarksnumbered=true, bookmarksopen=true, colorlinks=true, pdfborder=001, citecolor=blue, linkcolor=blue, linktocpage=true] {hyperref}

\begin{document}

\preprint{APS/123-QED}

\title{Magnon boundary states tailored by longitudinal spin-spin interactions and topology}

\author{Wenjie Liu$^{1,2,3}$}

\author{Yongguan Ke$^{1,2,3}$}
\altaffiliation{Email: keyg@mail2.sysu.edu.cn}

\author{Zhoutao Lei$^{2,3}$}

\author{Chaohong Lee$^{1,2,3}$}
\altaffiliation{Email: chleecn@szu.edu.cn}

\affiliation{$^{1}$College of Physics and Optoelectronic Engineering, Shenzhen University, Shenzhen 518060, China}

\affiliation{$^{2}$Guangdong Provincial Key Laboratory of Quantum Metrology and Sensing $\&$ School of Physics and Astronomy, Sun Yat-Sen University (Zhuhai Campus), Zhuhai 519082, China}

\affiliation{$^{3}$State Key Laboratory of Optoelectronic Materials and Technologies, Sun Yat-Sen University (Guangzhou Campus), Guangzhou 510275, China}

\date{\today}


\begin{abstract}
Since longitudinal spin-spin interaction is ubiquitous in magnetic materials, it is very interesting to explore the interplay between topology and longitudinal spin-spin interaction.
Here, we examine the role of longitudinal spin-spin interaction on topological magnon excitations.
Remarkably, even for single-magnon excitations, we discover topological edge states and defect edge states of magnon excitations in a dimerized Heisenberg XXZ chain and their topological properties can be distinguished via adiabatic quantum transport.
We uncover topological phase transitions induced by longitudinal spin-spin interactions whose boundary is analytically obtained via the transfer matrix method.
For multi-magnon excitations, even-magnon bound states are found to be always topologically trivial, but odd-magnon bound states may be topologically nontrivial due to the interplay between the transverse dimerization and the longitudinal spin-spin interaction.
For two-dimensional spin systems, the longitudinal spin-spin interaction contributes to the coexistence of defect corner states, second-order topological corner states and first-order topological edge states.
Our work opens an avenue for exploring topological magnon excitations and has potential applications in topological magnon devices.
\end{abstract}
	
\maketitle
	
\section{Introduction}\label{introduction}

Topological band theory underpins fertile topological states of matter~\cite{HasanRevModPhys823045,QiRevModPhys8310572011}, ranging from conventional to higher-order topological insulators.
According to the bulk-boundary correspondence (BBC)~\cite{PhysRevB4811851,PhysRevLett713697}, topological invariants of bulk bands are intimately related to robust boundary states.
Taking the Su-Schrieffer-Hegger (SSH) model~\cite{SuWPPhysRevLett1979,PhysRevB101134423} as an example, the bulk topology characterized by a winding number $\nu$ corresponds to $\nu$ pairs of edge states under open boundary condition.
Another typical example is the Benalcazar-Bernevig-Hughes (BBH) model~\cite{2017Science35761,2017PhysRevB96245115}, in which the nontrivial quantized quadrupole moment indicates four corner states in two dimensions.
Not limited to fermionic particles, the above wisdom has been widely applied to study topological bosonic excitations such as magnons~\cite{PhysRevLett.90.167204,PhysRevB.96.224414,Malki2020,ZXLiPhysRep2021,Bonbien2021,PhysRevX11021061}, phonons~\cite{ZhangPhysRevLett105225901} and photons~\cite{PhysRevLett100013904}.
	
Topological magnons, a kind of collective excitations over trivial ground states of magnetic materials, have provided new insights into topological states and potential applications such as topological magnon laser~\cite{GalHarariScience2018,MiguelScience2018}, magnon spintronics~\cite{ChumakNatPhy2015,DanielNatC2019} and topological magnetic memory~\cite{ARMellnikNature2014,PhysRevLett119077702}.
Parallel to the electronic counterpart, magnon Hall effect was theoretically predicted and experimentally observed~\cite{ScienceYOnose2010}, and Dirac magnons~\cite{PhysRevB.94.075401,PhysRevLett.119.247202,PhysRevX.8.011010,SBaoNatureComm2018,PhysRevX.10.011062}, Weyl magnons~\cite{FYLiNatureComm2016,PhysRevLett.117.157204,PhysRevB.95.224403,PhysRevB.96.104437,PhysRevB.97.115162,PhysRevB.99.214413,PhysRevResearch.2.013063}, nodal line magnons~\cite{PhysRevB.95.014418,PhysRevResearch.2.023282}, topological magnon polarons~\cite{PhysRevLett117217205,PhysRevLett123167202,PhysRevLett123237207,PhysRevLett124147204}, higher-order topological magnons~\cite{ASilJPCM2020,PhysRevB104024406} were theoretically predicted.

Because it is generally difficult to measure bulk topology of magnons, according to the BBC, edge states have been utilized as an experimental signature.
Most of the topological magnon excitations previously focused on are the isotropic Heisenberg interaction, where the role of the longitudinal spin-spin interaction is weak enough to be neglected~\cite{ScienceYOnose2010,2013PhysRevB87144101,2014NatCommun54815,2014PhysRevB90024412,2015PhysRevLett115147201,2016PhysRevLett117227201,2018PhysRevX8041028,2018PhysRevB97081106,2021PhysRevX11021061,PhysRevLett.117.157204,FYLiNatureComm2016,PhysRevX.8.011010,ASilJPCM2020,PhysRevB104024406}.
There have emerged many interesting topological boundary states with nontrivial magnon bands, such as  chiral magnon edge states~\cite{ScienceYOnose2010,2013PhysRevB87144101,2014NatCommun54815,2014PhysRevB90024412,2015PhysRevLett115147201,2016PhysRevLett117227201,2018PhysRevX8041028,2018PhysRevB97081106,2021PhysRevX11021061}, magnon surface states of Dirac and Weyl magnons~\cite{PhysRevLett.117.157204,FYLiNatureComm2016,PhysRevX.8.011010}, and topological magnon corner states~\cite{ASilJPCM2020,PhysRevB104024406}.
The conventional BBC breaks down in various situations, including non-Hermitian topological systems with the skin effect~\cite{2018PhysRevLett121086803,2018PhysRevLett121136802,2019PhysRevB99075130} and defect edge states~\cite{2016PhysRevLett116133903,2020PhysRevB101121116,2020IntJModPhysB}, hole-edge states in spinless SSH model with interaction~\cite{2017PhysRevB95115443}, two-dimensional topological insulators with the coexistence isolated corner states and gapless edge states~\cite{2109arXiv}, Floquet topological systems with anomalous edge modes~\cite{2013PhysRevX3031005}, and continuous models with ghost edge modes~\cite{2020PhysRevResearch2013147}.
Although the intrinsic longitudinal spin-spin interaction is ubiquitous,
it is still unclear {\em how the longitudinal spin-spin interactions affects topological magnon excitations in spin systems}.

In this article, we reveal the effect of longitudinal spin-spin interactions on magnon boundary states both in one- and two-dimensional topological spin systems as variants of SSH and BBH models, respectively.
Even for single-magnon excitations, topological edge states (mainly distributed at the second or the next to last site) and defect edge states (mainly distributed at the first or the last site) may coexist in a dimerized spin chain.
This is because the strong longitudinal spin-spin interaction gives rise to defects at boundaries, and make inter-cell and intra-cell couplings swap their positions.
By analyzing the variation of energy spectrum, spin magnetization and adiabatical topological pumping, two types of edge states are well distinguished.
The staggered magnetization is proposed to identify the transitions from topological trivial (nontrivial) phase to nontrivial (trivial) one in the parameter space spanned by longitudinal spin-spin interaction and transverse dimerization.
Furthermore, we find the topology of bound magnons sensitively depends on the number of magnon excitations, which is termed as even-odd effect.
Remarkably, besides to defect corner states and topological sub-corner states, there have first-order topological edge states in two-dimensional BBH-type spin systems which is understood via a second-order tunneling process.
Such three types of boundary states following distinct adiabatic pumping provide alternative ways for designing spin devices.

This paper is organized as follows.
In Sec.~\ref{onedimensional}, we examine the effect of longitudinal spin-spin interaction on a dimerized XXZ chain for topological single-magnon excitations in Sec.~\ref{topologicalsinglemagnon} and  topological multi-magnon excitations in Sec.~\ref{topologicalmultimagnon}.
Spin dynamics is used to detect the magnon-excitation states in Sec.~\ref{appendixDet}.
In Sec.~\ref{appendixHOMI}, we analyze the longitudinal spin-spin interaction in a two-dimensional BBH-type spin systems.
In Sec.~\ref{conclusiondicussion}, we give a conclusion and discussion.

\section{One-dimensional dimerized XXZ chains} \label{onedimensional}

We first consider an open Heisenberg XXZ chain with transverse dimerization,
\begin{equation}\label{DimerizedSpinChain}
	\hat{H}=-\sum_{l}^{}\left\{\left[\left(J+(-1)^{l} \delta_0\right)\hat{S}^+_l\hat{S}^-_{l+1}+\mathrm{H.c.}\right]+\Delta \hat{S}_{l}^{z} \hat{S}_{l+1}^{z} \right\}
\end{equation}
with the lattice index $l$, the spin-$\frac{1}{2}$ operators $\hat{S}^{(x,y,z)}_l$, and the spin-raising (-lowing) operators $\hat{S}^{\pm}_l=\hat{S}^x_l\pm i\hat{S}^y_l$.
$J\pm \delta_0$ respectively denote the inter-cell and intra-cell spin-exchange strengths, and $\Delta$ is the longitudinal spin-spin interaction.
Without loss of generality, we set $J=1$ and $\Delta>0$.
For a large $\Delta$, all spins downward $|\downarrow \downarrow \cdots \downarrow\rangle$ is a ferromagnetic ground state, in which flipping a spin upward creates a magnon excitation.
In contrast to studies of topological ground states~\cite{PhysRevB81014505,PhysRevLett1100753032013,PhysRevLett1102153012013,PhysRevLett1102604052013,PhysRevB93155133,PhysRevA990121222019,PhysRevLett1280434022022}, topological magnon excitations are associated with excited states.
Because $[\hat{H}, \sum_l\hat{S}_{l}^{z}]=0$, the subspaces with different magnon numbers are decoupled.
Below we study a $2L$-site spin chain of $L$ two-site cells, except the adiabatic topological pumping of edge states (where the site number is odd).

Viewing magnons as quasiparticles, the spin exchange and longitudinal spin-spin interaction correspond to nearest-neighbor hopping and magnon-magnon interaction, respectively.
At the first glance, since the magnon-magnon interaction is absent in single-magnon excitations, the topology of single-magnon excitations was supposed to behave as the celebrated SSH model~\cite{SuWPPhysRevLett1979,PhysRevB101134423}.
In such a SSH model, the bulk topology is characterized by the winding number $\nu$, where $\nu=1$ for $\delta_0>0$ and $\nu=0$ for $\delta_0 <0$.
The conventional BBC indicates that a pair of edge states do (not) exist when the intra-cell spin exchange is weaker (stronger) than the inter-cell spin exchange.
However, even for single-magnon excitations, we find that the conventional BBC becomes invalid in our system, in which the longitudinal spin-spin interaction still plays a crucial role.
	

\subsection{Topological single-magnon excitations} \label{topologicalsinglemagnon}

\begin{figure}[htp]
\center
\includegraphics[width=0.5\textwidth]{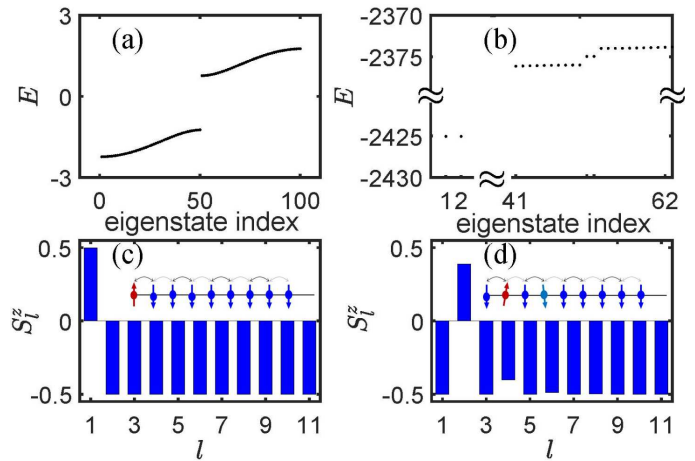}
\caption{(Color online) Single-magnon excitations. (a) and (b): single-magnon energy spectra respectively with $\Delta=0.01$ and $\Delta=100$.
The eigenstate index is ordered for increasing values of the energy.
The other parameters are chosen as $J=1$, $\delta_{0}=-0.5$ and $L=50$.
(c) and (d): spin magnetization distributions $S_{l}^{z}=\langle\Psi|\hat{S}_{l}^{z}| \Psi\rangle$ for the isolated states far from the bottom band and lying in the energy gap in (b).
The insets respectively correspond to the sketches of edge states.
\label{fig:onemagnonspectrum}}
\end{figure}

Below we discuss how longitudinal spin-spin interaction affects topological edge states in single-magnon systems, which is ignored in the observation of topological magnon insulator states~\cite{CaiPhysRevLett1230805012019}.
	%
We show single-magnon energy spectra for weak and strong longitudinal spin-spin interactions in Figs.~\ref{fig:onemagnonspectrum}(a) and~\ref{fig:onemagnonspectrum}(b), respectively.
For $\Delta=0.01$, there are only two separated energy bands, similar to the trivial SSH model.
However, for $\Delta=100$, four additional isolated edge states appear:
the two below the bottom band (the first two states) correspond to a single magnon strongly confined at the leftmost or rightmost site,
and the other two in the band gap (the ${(L+1)}$-th and ${(L+2)}$-th states) correspond to a single magnon strongly localized at the second or the next to last site;
see the spin magnetization $S_{l}^{z}=\langle\Psi|\hat{S}_{l}^{z}| \Psi\rangle$ in Figs.~\ref{fig:onemagnonspectrum}(c) and (d), respectively.
The insets in Figs.~\ref{fig:onemagnonspectrum}(c) and (d) schematically display the two types of edge states at left, while their degenerate counterparts are symmetrically localized at right.
Surprisingly, the edge states appear in the system of strong longitudinal spin-spin interaction even when the bulk topology was supposed to be trivial with $\delta_0<0$.
A question naturally arises: {\em what is the emergence mechanism of these edge states?}

\begin{figure}[htp]
\center
\includegraphics[width=0.5\textwidth]{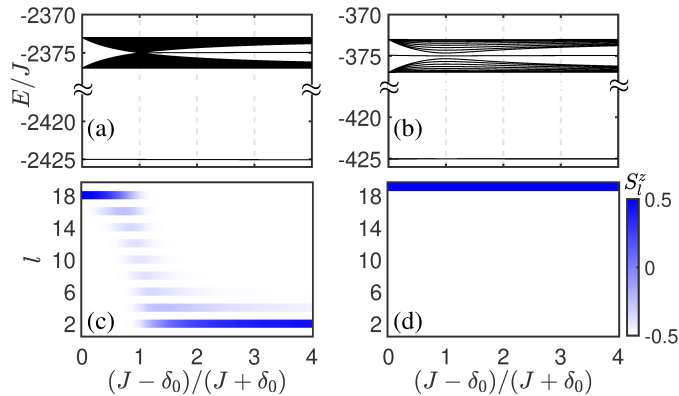}
\caption{(Color online) Appearance of topological edge states.
(a) Single-magnon excitation energy versus the transverse spin-exchange ratio $(J-\delta_{0})/(J+\delta_{0})$ for longitudinal spin-spin interaction $\Delta=100$ with $L=50$.
The energy spectrum (b) and adiabatic topological pumping of topological (c) and defect (d) edge states for a 19-site system with $\Delta=100$.
The other parameter is $J=1$.
\label{fig:topogicalinvariant}}
\end{figure}

To understand the origin of these edge states, we analyze their behaviors across the topological transition point $(J-\delta_{0})/ (J+\delta_{0})=1$ of the conventional SSH model.
When the longitudinal spin-spin interaction is weak enough, similar to a conventional SSH model, edge states appear in the band gap when $\delta_{0}>0$, which corresponds to nontrivial topology of the bulk band.
However, for strong enough $\Delta$, the edge states in the band gap appear when $\delta_0<0$; see Fig.~\ref{fig:topogicalinvariant}(a) with $\Delta=100$.
These edge states in the band gap indicate that the conventional BBC is broken by the strong longitudinal spin-spin interaction.
Besides, the edge states below the bottom band always exist regardless of the dimerization strength $\delta_0$.
This means that the edge states below the bottom band are not related to topology but are solely induced by the longitudinal spin-spin interaction.

To uncover these mysterious edge states, we revisit the mapping from longitudinal spin-spin interaction to magnon-magnon interaction.
From the perspective of magnon excitations under open boundaries, we discover that intrinsic longitudinal spin-spin interaction has two main effects: one is defect potential at the end points, and the other is magnon-magnon interaction.
Taking the open three-spin chain as an example, single-magnon non-edge states $(\left|\downarrow \uparrow \downarrow\right\rangle)$ and the edge states $(\left|\uparrow \downarrow \downarrow\right\rangle, \left|\downarrow \downarrow \uparrow\right\rangle)$ have onsite energies $\Delta/4 \pm \Delta/4$, respectively.
The longitudinal interaction contributes an energy offset $\Delta/2$ between the magnon at the edge sites and the other bulk sites~\cite{LiuPhysRevA2021}.
This means that the cooperation between the open boundary condition and the longitudinal spin-spin interaction induces effective on-site defects at the edge sites.
A large $\Delta/2$ will trap a magnon at an end point to form the non-topological edge state, which can be dubbed as defect edge state.
In the limit of $\Delta\rightarrow \infty$, the first and the last sites are decoupled from other bulk sites and can then be understood as a new interface.
Thus the second and the next to last sites act as a new ``boundary'' of a topological SSH lattice, in which the intra- and inter-cell spin exchanges are swapped.
Consequently, the topology of the renormalized SSH model is opposite to the original trivial one, and topological edge states appear at the second and the next to last sites.
Hence, these sub-edge states in the band gap belong to topological edge states.

To further distinguish two types of edge states, we adiabatically sweep $(J-\delta_0)/(J+\delta_0)$ across the phase transition point in a 19-site system with $\Delta=100$.
Its instantaneous energy spectrum is shown in Fig.~\ref{fig:topogicalinvariant}(b).
With the two types of edge states for $(J-\delta_0)/(J+\delta_0)=0$ as initial states, after tracking the change of spin magnetization we find that the topological edge state can be adiabatically transferred from one end to the other end [see Fig.~\ref{fig:topogicalinvariant}(c)], while the defect edge state remains unchanged [see Fig.~\ref{fig:topogicalinvariant}(d)].
This means that observation of edge states is not enough to support bulk-boundary correspondence and that adiabatic topological pumping of edge states is a more rigorous method to demonstrate the bulk topology.

\begin{figure}[htp]
\center
\includegraphics[width=0.5\textwidth]{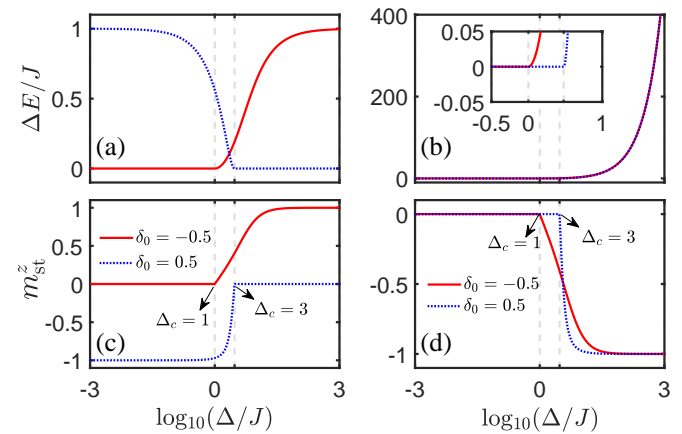}
\caption{(Color online) The band-edge gap $\Delta E$ as a function of the longitudinal spin-spin interaction $\Delta$: (a) topological edge state and (b) non-topological edge state.
The staggered magnetization $m_{\mathrm{st}}^{z}$ as a function of the longitudinal spin-spin interaction $\Delta$ for (c) the ($L+1$)-th state and (d) the first state.
%
The inset of (b) zooms in its region around critical points.
Our calculations are performed with $\delta_{0}=-0.5$ (red solid lines) and $\delta_{0}=0.5$ (blue dotted lines) and $L=5000$.
\label{fig:chosendelta0}}
\end{figure}

Owing to the longitudinal spin-spin interactions, there exist transitions between topologically trivial and nontrivial phases as $\Delta$ increases.
We investigate the band-edge gap $\Delta E$ to determine the critical point $\Delta_c$, both analytically and numerically.
For topological edge state, in the thermodynamic limit $\Delta E$ naturally tends to zero in the topologically trivial phase, and takes a finite value due to the appearance of topological edge states in the topologically nontrivial phase.
We calculate its band-edge gap as a function of $\Delta$ for $(J-\delta_{0})/ (J+\delta_{0})=3$ with $\delta_0=-0.5$ (red solid line) and $(J-\delta_{0})/ (J+\delta_{0})=1/3$ with $\delta_0=0.5$ (blue dotted line) for a lattice size $L=5000$; see Fig.~\ref{fig:chosendelta0}(a).
The band-edge gap can successfully identify the critical points $\Delta_c$, no matter for the increase from zero to finite value or vice versa.
The corresponding $m_{\mathrm{st}}^{z}=\sum_{l=1}^{2L}(-1)^{l}\langle\Psi|\hat{S}_{l}^{z}|\Psi\rangle$ of the ($L+1$)-th state is exhibited in Fig.~\ref{fig:chosendelta0}(c).
There are two degenerate topological edge states in topologically nontrivial phases where the ($L+1$)-th state is used to represent the left topological edge state.
$m_{\mathrm{st}}^{z}\approx-1$ means the corresponding left topological edge state almost distributes at odd sites where a single-magnon excitation mainly locates at the first site and decays at other odd sites.
However, differently, if left topological edge state mainly distributes at even sites, it has $m_{\mathrm{st}}^{z}\approx1$.

For non-topological edge state, $\Delta E=E_3-E_2$ represents the energy gap between the second and third eigenstates, which remains vanishingly small when $\Delta<\Delta_c$.
Unlike the topological edge states finally lying in the middle of two bulk bands, the non-topological edge states linearly increase with $\Delta$ after separating from the bottom bands, as shown in Fig.~\ref{fig:chosendelta0}(b).
The staggered magnetization of the first state in Fig.~\ref{fig:chosendelta0}(d) reflects, after crossing the critical point $\Delta_c$, the left non-topological edge state considerably distributes at odd sites as $\Delta$ increases.
Two degenerate non-topological edge states always appear as a pair where the first state is used to represent the left non-topological edge state.
The transition point induced by longitudinal spin-spin interaction depends on the appearance or disappearance of non-topological states.
Importantly, using the transfer matrix method, the condition for the appearance of the defect edge states can be analytically given by $\Delta>\Delta_c$ with $\Delta_c=2(J+\delta_0)$ (see Appendix~\ref{appendixCritical} for details).
The analytical critical points $\Delta_c=1$ and $\Delta_c=3$ for two cases are added in Fig.~\ref{fig:chosendelta0} with gray dashed lines which are well consistent with the numerical ones.

\begin{figure}[htp]
\center
\includegraphics[width=0.38\textwidth]{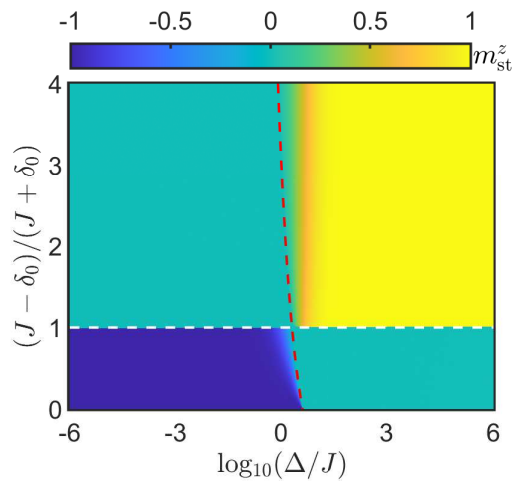}
\caption{(Color online) The staggered magnetization $m_{\mathrm{st}}^{z}$ of ($L+1$)-th state as a function of the longitudinal spin-spin interaction $\Delta$ and the transverse dimerization $(J-\delta_{0})/ (J+\delta_{0})$ for a $2L$-site system with $L=1000$.
Gray shading is a region where the staggered magnetization cannot be integer owing to the absence of chiral symmetry.
The critical lines $\Delta_c=2(J+\delta_0)$ and $(J+\delta_0)/(J-\delta_0)=1$ are marked by the red and white dashed lines, respectively.
The other parameter is $J=1$.
\label{fig:windingnumber}}
\end{figure}

Because the longitudinal spin-spin interaction breaks the chiral symmetry and invalidates topological invariant, we calculate the staggered magnetization to witness the transition from topologically trivial to nontrivial phases.
Fig.~\ref{fig:windingnumber} manifests the staggered magnetization $m_{\mathrm{st}}^{z}$ of ($L+1$)-th state in the parameter space spanned by $\Delta$ and $(J-\delta_{0})/ (J+\delta_{0})$.
Although $m_{\mathrm{st}}^{z}$ vanishes in topologically trivial phases, it takes a finite negative (positive) value in the left-bottom (right-top) region.
For a weak longitudinal spin-spin interaction, the topological characterization behaves like the conventional SSH model: topologically nontrivial phase for $(J-\delta_{0})/ (J+\delta_{0})<1$ while for $(J-\delta_{0})/ (J+\delta_{0})>1$ it turns into the trivial one.
Tuning the longitudinal spin-spin interaction strong enough accompanied with the appearance of the non-topological edge states, topological edge states exist for $(J-\delta_{0})/ (J+\delta_{0})>1$ while disappear for $(J-\delta_{0})/ (J+\delta_{0})<1$.
The analytical critical lines $\Delta_c=2(J+\delta_0)$ and $(J-\delta_{0})/ (J+\delta_{0})=1$ are respectively added in Fig.~\ref{fig:windingnumber} with red and white dashed lines,
indicating that the analytical results agree well with the numerical ones.
Similar to the nonlinear Thouless pumping~\cite{2021Nature59663}, near the critical line $\Delta_c$ there exists a narrow region (the gray region) in which $0<|m_{\mathrm{st}}^{z}|<1$.
From the perspective of distribution properties of topological edge states, such a staggered magnetization can serve as an effective topological indicator to understand the topological phase transitions.

\subsection{Topological multi-magnon excitations} \label{topologicalmultimagnon}

The longitudinal spin-spin interaction provides not only boundary defects but also nearest-neighbor magnon-magnon interaction in multi-magnon excitations.
The magnon-magnon interaction can bind magnons together as bound states~\cite{2012PhysRevLett108077206,2013Nature50276,2014PhysRevLett112257204,2017PhysRevB96195134}.
When the longitudinal spin-spin interaction is sufficiently strong,
one can treat the transverse term as a perturbation to the longitudinal one to analytically derive an effective model for well explaining the bound states.
Here, we just present results for multi-magnon bound states (see Appendix~\ref{appendixA} and Appendix~\ref{appendixB} for unbound magnons).
The bound states can be treated as a quasiparticle that is governed by an effective Hamiltonian.
Taking two and three magnons as examples, the effective hopping strengths of bound states in the bulk are given by $J_{\mathrm{Eff}}^{(2)}={\left(J+\delta_{0}\right)\left(J-\delta_{0}\right)}/{\Delta}$ and $J_{\mathrm{Eff}}^{(3)}={\left(J-\delta_{0}\right)\left(J+\delta_{0}\right)\left[J+\delta_{0}(-1)^l\right]}/{\Delta^2}$
(see Appendix~\ref{appendixA} and Appendix~\ref{appendixB} for details).
Because the effective hopping strengths of two- and three-magnon bound states are respectively uniform and site-dependent,
we know that two-magnon bound states are topologically trivial and three-magnon bound states may inherit topology from the SSH model.
Due to no more than one magnon at one site, the effective hopping originates from the $n$-th order process for $n$-magnon excitations.
From the effective $n$-th order hopping process, we find that the corresponding effective hopping strength is proportional to $[J+\delta_0(-1)^l]^{mod(n,2)}(J+\delta_0)^{\lfloor n/2 \rfloor}(J-\delta_0)^{\lfloor n/2 \rfloor}$, where $\lfloor{x\rfloor}$ takes the closest integer value less than or equal to $x$.
We generalize this result to an even-odd effect, that is, even-magnon bound states are trivial and odd-magnon bound states may have nontrivial topology.

\begin{figure}[htp]
\center
\includegraphics[width=0.45\textwidth]{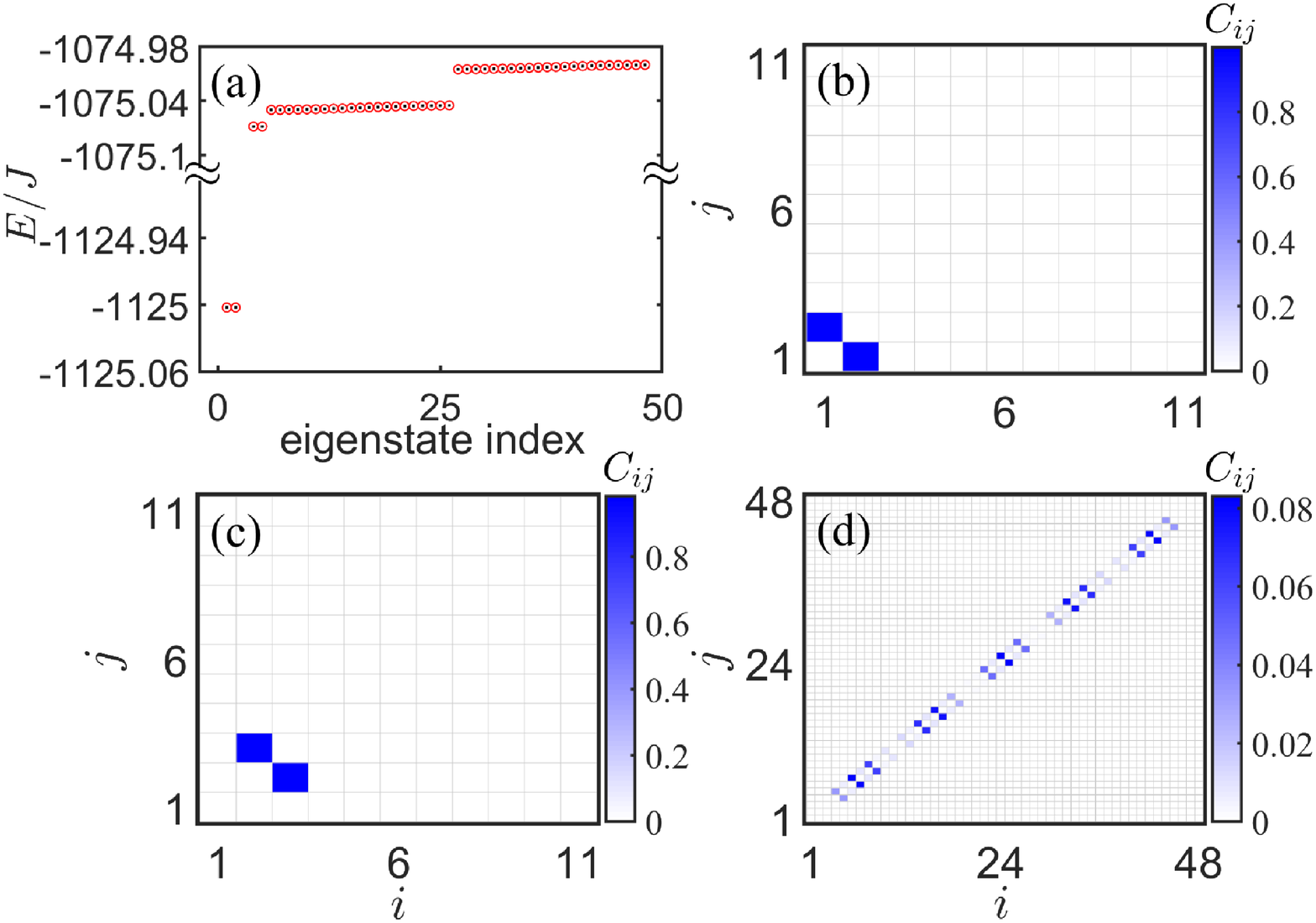}
\caption{(Color online) Two-magnon bound-state subspace for a strong longitudinal spin-spin interaction ($\Delta=100$).
(a) Bound-state energy spectrum in ascending order for the values of the energies.
The black dots denote the energies regarding the Hamiltonian~\eqref{DimerizedSpinChain},
and the red circles are the energies obtained by the effective model $\hat{H}_{\mathrm{Eff}}^{(2)}$~\eqref{Heff2}.
The parameters are chosen as $J=1$, $\delta_{0}=-0.5$ and $L=24$.
(b)-(d) are respectively two-magnon correlation distributions of chosen eigenstates in (a).
\label{fig:TwoBoundMagnons}}
\end{figure}

We further analyze topological states in the two- and three-magnon bound-state subspaces.
For a strong longitudinal interaction, we obtain the bound-state subspace consisting of two-magnon bound states in Fig.~\ref{fig:TwoBoundMagnons}(a).
The eigenstate index on the horizontal axis is ordered with increasing values of the energy.
The parameters are chosen as $J=1$, $\delta_{0}=-0.5$, $\Delta=100$ and $L=24$.
The energy spectrum of the effective model $\hat{H}_{\mathrm{Eff}}^{(2)}$ is added in Fig.~\ref{fig:TwoBoundMagnons}(a) with red circles.
It can be observed that the numerical results (black dots) from the Hamiltonian~\eqref{DimerizedSpinChain} perfectly agree with the analytical ones (red circles) given by the effective model $\hat{H}_{\mathrm{Eff}}^{(2)}$~\eqref{Heff2} whose validity has been analyzed in Fig.~\ref{fig:ValidityTwoMagnons}.
The bound-state subspace contains four isolated bands which are ordered for increasing values of the energy.
We extract one state of the first band to calculate the two-magnon correlation function $C_{i j}=\langle\Psi|\hat{S}_i^{+} \hat{S}_j^{+} \hat{S}_j^{-} \hat{S}_i^{-}| \Psi\rangle$ in Fig.~\ref{fig:TwoBoundMagnons}(b), which indicates eigenstates with two magnons bound at one end point as a two-magnon bound edge state.
The correlation properties of the second-band states show that the bound magnons are mostly distributed at one sub-edge to form a two-magnon bound sub-edge state [Fig.~\ref{fig:TwoBoundMagnons}(c)].
The two remaining continuum bands include eigenstates with two magnons as bound pairs [Fig.~\ref{fig:TwoBoundMagnons}(d)].
The emergence of bound edge states is due to the fact that the longitudinal spin-spin interaction creates potentials trapping one magnon at one end point and binding the other together.
The presence of two-magnon bound sub-edge states results from the effective potentials at the sub-edges (the second and the next to last sites) in the effective model $\hat{H}_{\mathrm{Eff}}^{(2)}$~\eqref{Heff2}.

\begin{figure}[htp]
\center
\includegraphics[width=0.45\textwidth]{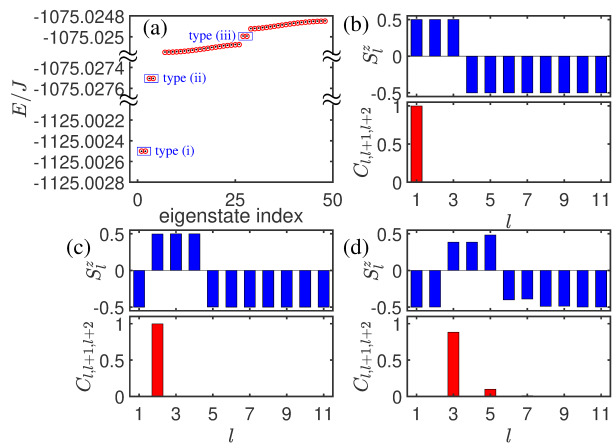}
\caption{(Color online) Three-magnon bound-state subspace for a strong longitudinal spin-spin interaction ($\Delta=100$). (a) Bound-state energy spectrum as a function of eigenstate index in the ascending order of energy.
	The black dots denote the eigen-energies of the Hamiltonian~\eqref{DimerizedSpinChain}, and the red circles are the eigen-energies of the effective Hamiltonian $\hat{H}_{\mathrm{Eff}}^{(3)}$~\eqref{Heff3}.
	(b)-(d): spin magnetization distributions $S_{l}^{z}$ and three-magnon correlations $C_{l,l+1,l+2}$ for three types of three-magnon bound edge states in (a).
	Here, $\sum_{l=1}^{2L-2} C_{l, l+1, l+2} \geqslant 0.9997$.
	The parameters are chosen as $J=1$, $\delta_{0}=0.5$ and $L=24$.
\label{fig:Threemagnons}}
\end{figure}

Fig.~\ref{fig:Threemagnons}(a) shows the energy spectrum for three-magnon bound states.
The energy spectrum of the effective model $\hat{H}_{\mathrm{Eff}}^{(3)}$~\eqref{Heff3} is added in Fig.~\ref{fig:Threemagnons}(a) with red circles.
It can be observed that the numerical results (black dots) from the Hamiltonian~\eqref{DimerizedSpinChain} agree well with the analytical ones (red circles) given by the effective model $\hat{H}_{\mathrm{Eff}}^{(3)}$~\eqref{Heff3} whose validity has been analyzed in Fig.~\ref{fig:ValidityThreeMagnons}.
There are three types of three-magnon bound edge states: (i) bounded to the first site, (ii) bounded to the second site, and (iii)  bounded to the third site with energy in the band gap, whose spin magnetizations and three-magnon correlations are shown in Fig.~\ref{fig:Threemagnons}(b, c, d), respectively.
Here, the three-magnon correlation is defined as $C_{l,l+1,l+2}=\langle\Psi|\hat{S}_{l}^{+} \hat{S}_{l+1}^{+} \hat{S}_{l+2}^{+}\hat{S}_{l+2}^{-}\hat{S}_{l+1}^{-} \hat{S}_{l}^{-}| \Psi\rangle$.
The type-(i,ii) three-magnon bound states are attributed to the emergent defects at the first and second sites, while the type-(iii) three-magnon bound states are due to nontrivial topology of the bulk band (see the effective Hamiltonian $\hat{H}_{\mathrm{Eff}}^{(3)}$~\eqref{Heff3} in Appendix~\ref{appendixB} for details).
	
\subsection{Detecting magnon-excitation states via spin dynamics} \label{appendixDet}

\begin{figure*}[htp]
\center
\includegraphics[width=0.9\textwidth]{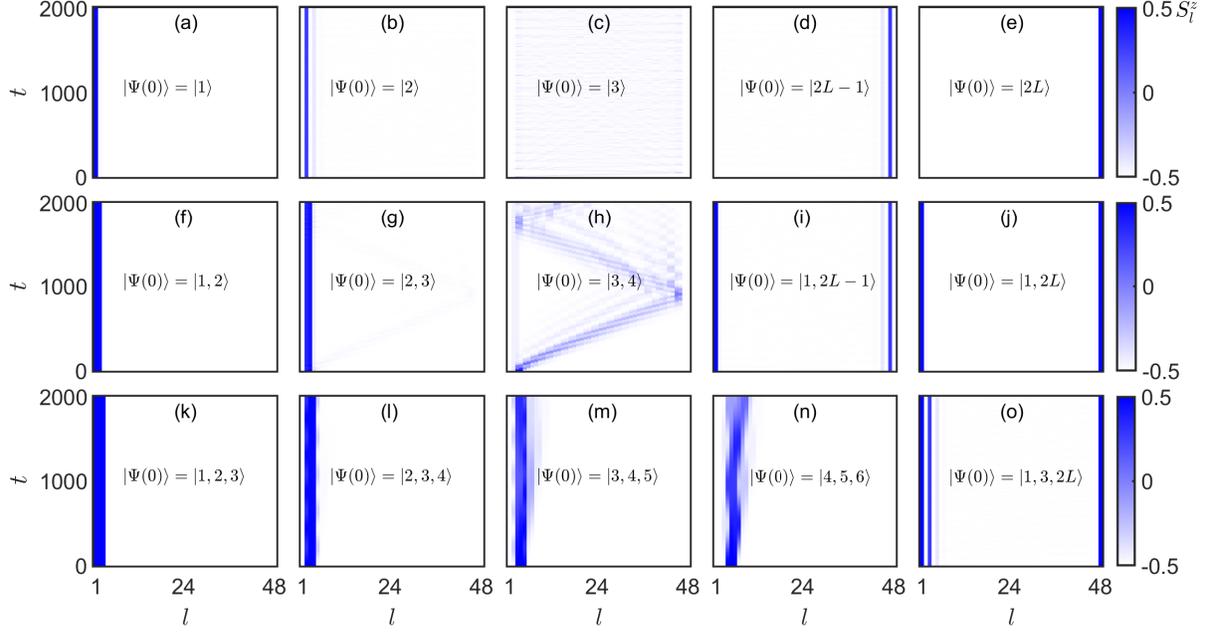}
\caption{(Color online) The time evolution of the spin magnetization distribution $S_{l}^{z}(t)=\langle\Psi(t)|\hat{S}_{l}^{z}|\Psi(t)\rangle$ obtained using TEBD algorithm.
The top, middle, and bottom rows correspond to single-magnon excitations $N_m=1$ with $\delta_0=-0.5$, two-magnon excitations $N_m=2$ with $\delta_0=-0.5$, and
three-magnon excitations $N_m=3$ with $\delta_0=0.5$, respectively.
For all three rows, the chosen initial states are shown from left to right, respectively.
The evolved time is set as $t=2000$.
The other parameters are chosen as $J=1$ and $\Delta=100$. The system size is $2L=48$.
\label{fig:spindynamics}}
\end{figure*}

Spin dynamics provides new insights into the detection of magnon-excitation states.
Referring to the above analysis, a rich variety of magnon-excitation states appear as tuning intrinsic system parameters and magnon-excitation numbers, especially for topological magnon edge states.
Topologically protected edge states have been employed to implement quantum state transfer~\cite{2012PhysRevLett109106402,2013NatCommun41585,2017QuantumSciTechnol2015001,2018PhysRevA98012331,2019PhysRevLett123034301}, disorder-immune photonic and phononic transport~\cite{2014NatPhotonics8821,2016NatPhys12621,2019RevModPhys91015006}, topological quantum computation~\cite{2015RevModPhys87137}, and topological laser~\cite{GalHarariScience2018,MiguelScience2018}.
We calculate the time evolution of the spin magnetization distribution $S_{l}^{z}(t)=\langle\Psi(t)|\hat{S}_{l}^{z}|\Psi(t)\rangle$ by using the time-evolving block decimation
(TEBD) algorithm~\cite{VidalPhysRevLett911479022003,VidalPhysRevLett930405022004}.
Initially, magnon excitations are located at different lattice sites, as shown in Fig.~\ref{fig:spindynamics} from $N_m=1$ to 3 magnons.
When initial magnon excitations are prepared with a high enough fidelity with the corresponding edge states, dynamical localization, where magnon excitations almost stay at the initial positions as time evolves, appears as a consequence of edge states.
The existence of non-topological edge states and topological edge state offers promising applications for manipulating the spin transports and designing the magnon spintronic devices.

\section{Two-dimensional BBH-type spin systems} \label{appendixHOMI}

\begin{figure*}[htp]
\center
\includegraphics[width=1\textwidth]{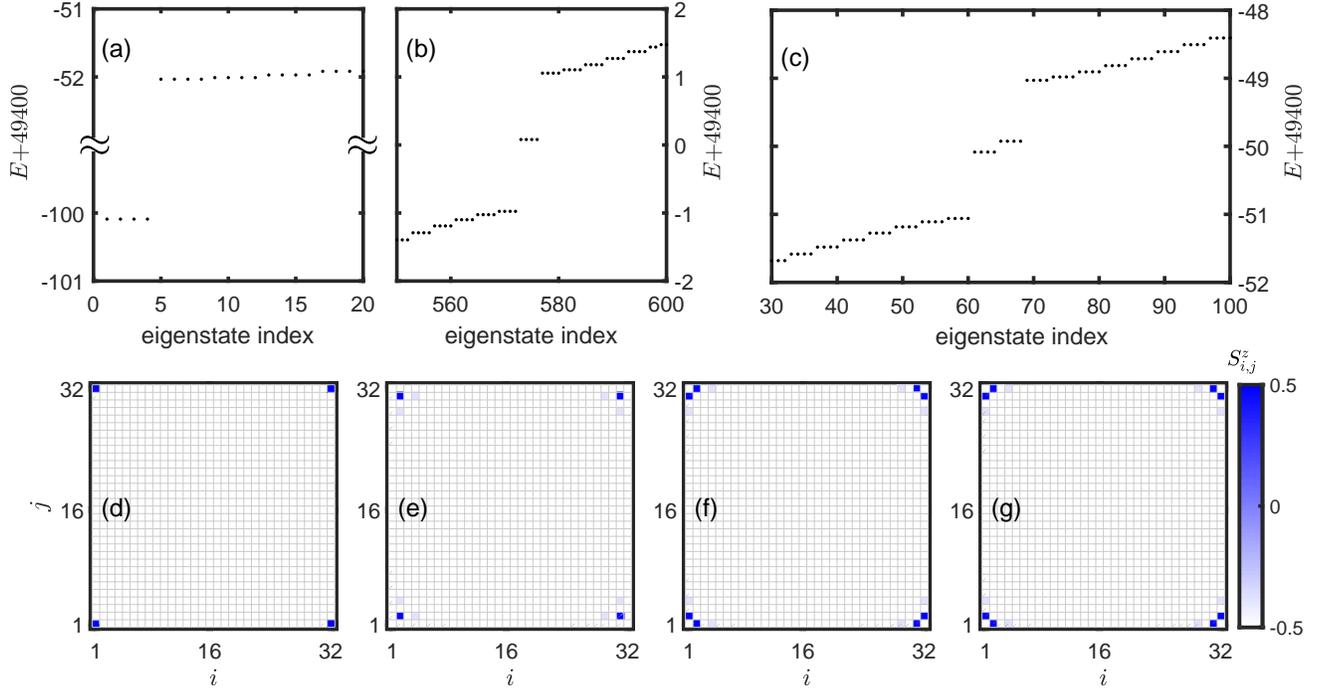}
\caption{(Color online) Single-magnon excitations in a two-dimensional BBH-type spin system.
Three parts of single-magnon excitation spectra: (a) defect corner states, (b) topological corner states as well as their neighboring eigenstates and (c) first-order topological edge states as well as their neighboring eigenstates.
The eigenstate index is ordered for increasing values of the energy.
Spin magnetization ${S}_{i,j}^{z}$ for defect corner states (d), topological corner states (e), lower (f) and upper four topological edge states (g).
The other parameters are chosen as $J=1$, $\delta_{0}=-0.5$ and $L=16$.
\label{fig:threeedgestates}}
\end{figure*}

After revealing the effects of intrinsic spin-spin interaction on magnon edge states in a dimerized spin chain,
it is reasonable to predict that the longitudinal spin-spin interaction plays a crucial role in topological magnon corner states by generating non-topological corner states.
Based on the quantized quadrupole moment in a Benalcazar-Bernevig-Hughes (BBH) model~\cite{2017Science35761,2017PhysRevB96245115}, our two-dimensional BBH-type spin system with longitudinal spin-spin interaction described by
\begin{eqnarray}\label{TDimerizedSpinChain}
\hat{H}&=&-\sum_{j=1}^{2 L} \sum_{i=1}^{2 L-1}\left[J+(-1)^{i} \delta_{0}\right] \hat{S}_{i, j}^{+} \hat{S}_{i+1, j}^{-}+\text {H.c. } \nonumber \\
&-& \sum_{i=1}^{2 L} \sum_{j=1}^{2 L-1}(-1)^i\left[J+(-1)^{j} \delta_{0}\right] \hat{S}_{i, j}^{+} \hat{S}_{i, j+1}^{-} +\text {H.c. }  \nonumber \\
&-&\Delta \hat{S}_{i, j}^{z} \hat{S}_{i+1, j}^{z}-\Delta \hat{S}_{i, j}^{z} \hat{S}_{i, j+1}^{z}.
\end{eqnarray}
%
We restrict to discuss the subspace of single-magnon excitations spanned by the basis $\left\{|i, j\rangle=\hat{S}_{i, j}^{+}|\downarrow \downarrow \downarrow \ldots \downarrow\rangle\right\}$ with $1 \leq i,j \leq 2 L$ and $|\downarrow \downarrow \ldots \downarrow\rangle$ being the ferromagnetic ground state.
When $\Delta=0$, the Hamiltonian~\eqref{TDimerizedSpinChain} for single-magnon excitations is equal to a BBH model~\cite{2017Science35761,2017PhysRevB96245115}, where two phases are distinct:
one is the topological phase that supports localized corner states almost distributing at four outmost corners when $\delta_0>0$, and the other is the trivial phase lacking the corner states when $\delta_0<0$.

As the longitudinal spin-spin interaction increases, the topological boundaries are gradually shifted one lattice site inward to support topological sub-corner states.
By analyzing the spin magnetization $S_{i, j}^z$ at position ($i$,$j$) for $\Delta=100$, four defect corner states occupy four outmost corners [see Fig.~\ref{fig:threeedgestates}(d)] away from the bulk bands [see Fig.~\ref{fig:threeedgestates}(a)].
In contrast to the quadrupolar topological insulator, four topological sub-corner states [see Fig.~\ref{fig:threeedgestates}(e)] are in the midgap of bulk bands [see Fig.~\ref{fig:threeedgestates}(b)] when $\delta_0<0$.
Unlike the absence of first-order topological edge states due to the vanishing dipole moment in BBH models, one can find that eight first-order topological edge states, lying inside the energy gap, are grouped into four degenerate pairs with energy splitting; see Fig.~\ref{fig:threeedgestates}(c).
Figs.~\ref{fig:threeedgestates}(f) and (g) respectively manifest the spin magnetization for the lower and upper four topological edge states in Fig.~\ref{fig:threeedgestates}(c).
The eigenstate index on the horizontal axis is ordered with increasing values of the energy.
The other parameters are chosen as $J=1$, $\delta_{0}=-0.5$ and $L=16$.
The emergence of first-order topological edge states is due to the fact that
the longitudinal spin-spin interaction creates potentials, making four boundary lines decouple to the bulk sites.
Each boundary line is equal to a one-dimensional dimerized spin chain that supports the coexistence of defect edge states and topological edge states.
Two defect edge states meeting at a corner create a defect corner state.
Eight first-order topological edge states originate from the four boundary lines.
However, the spin magnetization of first-order topological edge states in Figs.~\ref{fig:threeedgestates}(f) and (g)
is still unclear.

\begin{figure*}[htp]
	\center
	\includegraphics[width=1\textwidth]{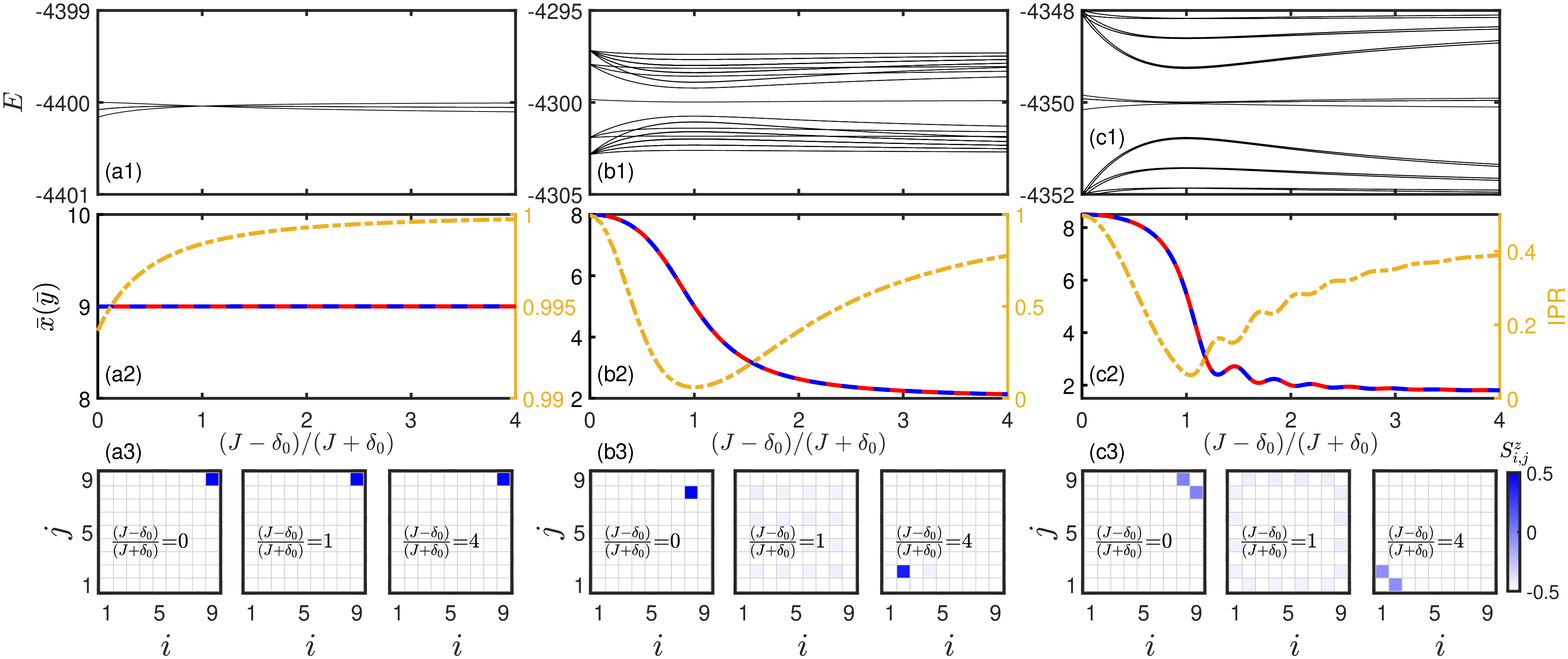}
	\caption{(Color online) Instantaneous energy spectra of adiabatic pumping for defect corner states (a1), topological corner state as well as its neighboring states (b1) and first-order topological edge states as well as their neighboring states (c1).
		(a2), (b2), (c2) Average positions $\bar{x}$ with red lines ($\bar{y}$ with blue dashed lines) in $x$ ($y$) direction and IPRs with yellow dashed-dotted lines of one of defect corner state, topological corner state and the lowest first-order topological edge state, respectively.
		(a3), (b3), (c3) The spin magnetizations at moments $(J-\delta_0)/(J+\delta_0)=0$,1,4 from left to right corresponding to (a2), (b2) and (c2), respectively.
		The other parameters are chosen as $J=1$, $\delta_{0}=-0.5$ and $L=5$.
		\label{fig:cornertransport}}
\end{figure*}

Without loss of generality, we focus on four positions [($1$,$1$), ($1$,$2$), ($2$,$1$), ($2$,$2$)] to explore such first-order topological edge states.
An energy difference $\Delta/2$ between the magnon at the edge sites and the other bulk sites comes from the longitudinal spin-spin interaction, which becomes $\Delta$ between the magnon at the corners and the other bulk sites.
Once ignoring an energy constant, the matrix of the above space spanned by the basis ($|1,1\rangle$, $|1,2\rangle$, $|2,1\rangle$, $|2,2\rangle$) is written as
\begin{equation}
\hat{H}_S=\left(\begin{array}{cccc}
\Delta & J-\delta_0 & -\left(J-\delta_0\right) & 0 \\
J-\delta_0 & \Delta / 2 & 0 & J-\delta_0 \\
-\left(J-\delta_0\right) & 0 & \Delta / 2 & J-\delta_0 \\
0 & J-\delta_0 & J-\delta_0 & 0
\end{array}\right).
\end{equation}
When the longitudinal spin-spin interaction is strong enough, the population in states $|1,1\rangle$ and $|2,2\rangle$ can be adiabatically eliminated.
Then the effective $2\times2$ model in basis ($|1,2\rangle$, $|2,1\rangle$) can be obtained as
\begin{equation}
	\left(\begin{array}{cc}0 & 4\left(J-\delta_0\right)^2 / \Delta \\ 4\left(J-\delta_0\right)^2 / \Delta & 0\end{array}\right).
\end{equation}
By solving the eigenequation, its eigenstates are $(|1,2\rangle \pm|2,1\rangle) / \sqrt{2}$ with eigenvalues $\pm 4\left(J-\delta_0\right)^2 / \Delta$, respectively.
The remaining six first-order topological edge states can be understood in the same way.
The topological edge states of four boundary lines hybridize through a second-order tunneling process to form the hybrid first-order topological edge states in Figs.~\ref{fig:threeedgestates}(f) and (g).

After removing two arrays from the $x$ and $y$ directions yielding a $(2L-1)$$\times$$(2L-1)$ square lattice, the adiabatic evolution is shown in Fig.~\ref{fig:cornertransport} with $L=5$.
The other parameters are chosen as $J=1$ and $\delta_{0}=-0.5$.
By adiabatically sweeping $(J-\delta_0)/(J+\delta_0)$, there are still four defect corner states with energy far away from the other states; see the instantaneous energy spectrum in Fig.~\ref{fig:cornertransport}(a1).
We choose one of four defect corner states as initial state, and the evolved state $|\psi(t)\rangle=\sum_{i,j} p_{i,j}(t) |i,j\rangle$ is governed by the
Schr\"{o}dinger equation $i \hbar \frac{\partial}{\partial t}|\psi(t)\rangle=\hat{H}|\psi(t)\rangle$.
The average position $\bar{x}=\sum_i i\left|p_{i,j}\right|^2$ ($\bar{y}=\sum_j j\left|p_{i,j}\right|^2$) of single-magnon excitations in the $x$ ($y$) direction is shown in Fig.~\ref{fig:cornertransport}(a2) with the red line and the blue dashed line, respectively.
It means the magnon excitation is well localized at the initial corner whose corresponding inverse participation ratios $\operatorname{IPR}=\sum_{i,j}\left|p_{i,j}\right|^4$ are shown in Fig.~\ref{fig:cornertransport}(a2) with the yellow dashed-dotted line for a localized state with a high value.
The spin magnetizations at moments $(J-\delta_0)/(J+\delta_0)=0$, $1$ and $4$ are also analyzed in Fig.~\ref{fig:cornertransport}(a3) from left to right, which clearly exhibits the unchanged corner state.
The other three defect corner states all behave like Figs.~\ref{fig:cornertransport}(a1), (a2) and (a3).

Fig.~\ref{fig:cornertransport}(b1) shows the instantaneous energy spectrum of topological corner state as well as its neighboring states.
Starting from the initial topological corner state, its average position and IPR are revealed in Fig.~\ref{fig:cornertransport}(b2).
It can be seen from Fig.~\ref{fig:cornertransport}(b3) from left to right that the initial state is set to be topological sub-corner state at the right-top sub-corner, then gradually spreads over the bulk sites around moment $(J-\delta_0)/(J+\delta_0)=1$, and finally transfers to be the topological sub-corner state at the left-bottom sub-corner for $(J-\delta_0)/(J+\delta_0)=4$.
In Fig.~\ref{fig:cornertransport}(c1), we plot the instantaneous energy spectrum of first-order topological edge states as well as their neighboring states.
It is found that four first-order topological edge states appear in the middle of two bands.
The energy splitting resulting from the second-order tunneling process allows the lowest and highest topological edge states to be separated from the middle two.
Specifically, we show the adiabatic transfer of the lowest first-order topological edge state in Fig.~\ref{fig:cornertransport}(c2).
The state transfer between two hybrid first-order topological edge states at diagonal lines is established through the extended edge states around the moment $(J-\delta_0)/(J+\delta_0)=1$; see Fig.~\ref{fig:cornertransport}(c3).

\section{Conclusion and discussion} \label{conclusiondicussion}

We elaborate the role of the longitudinal spin-spin interaction on topological magnon excitaitons ranging from
one-dimensional dimerized XXZ chains to two-dimensional BBH-type spin systems.
In a dimerized XXZ chain, we explore the appearance of topological edge states at the sub-edges, which can be distinguished by spin magnetization, variation of energy spectrum, and adiabatic topological pumping.
The topological phase transition induced by longitudinal spin-spin interaction in our system is accompanied by a bulk-edge gap closing of topological edge states rather than a bulk-band gap closing.
Its phase boundary is analytically derived via the transfer matrix method, and different phases are faithfully identified via the staggered magnetization.
We analytically obtain an effective model for revealing the even-odd effect of magnon-excitation number on topological magnons based upon the many-body degenerate perturbation theory.
Remarkably, the odd-even effect we find here is the first example in which the particle number serves as a degree of freedom to tune the topological properties.
Our results demonstrate that the magnon-magnon correlation of multi-magnon excitations plays a crucial role in topological multi-magnon states.
The interplay among longitudinal spin-spin interaction, transverse dimerization and magnon-magnon correlations leads to a rich variety of magnon excitations which are able to be flexibly detected via spin dynamics.
For a two-dimensional BBH-type spin system, the longitudinal spin-spin interaction is responsible for the coexistence of defect corner states, second-order topological corner states and first-order topological edge states.
Beyond the absence of first-order topological edge states in a typical BBH model, the hybrid first-order topological edge states are well explained by a second-order tunneling process.
These three types of magnon-excitation states with distinct adiabatic pumping offer alternative ways for state transfer.

We have demonstrated that longitudinal spin-spin interactions provide a versatile tool to engineer topological magnon states, which have been overlooked for a long time.
It deserves further study to generalize our method to other interacting systems such as long-range spin systems and extended Hubbard systems.

\begin{acknowledgments}
We acknowledge useful discussions with Linhu Li, Ling Lin and Li Zhang.
This work is supported by the National Natural Science Foundation of China (Grants No. 12025509, No. 11874434), the Key-Area Research and Development Program of GuangDong Province (Grants No. 2019B030330001), and the Science and Technology Program of Guangzhou (China) (Grants No. 201904020024).
W.L. is partially supported by the National Natural Science Foundation of China (Grant No. 12147108) and the
Fundamental Research Funds for the Central Universities, Sun Yat-Sen University (Grant No. 22qntd3101).
Y.K. is partially supported by the National Natural Science Foundation of China (Grant No. 11904419; No. 12275365).
\end{acknowledgments}

\appendix

\section{Derivation of the critical point} \label{appendixCritical}

Here we implement the transfer matrix approach~\cite{MacKinnonA1983,ChalkerJTPhysRevLett70982,RevModPhys69731} to derive the critical point $\Delta_c$.
For a $2L$-lattice system, the $N_m$-magnon Hilbert space can be spanned by the basis $\mathcal{B}^{(N_m)}=\left\{\left|l_{1} l_{2} \ldots l_{N_m}\right\rangle=\hat{S}_{l_{1}}^{+} \hat{S}_{l_{2}}^{+} \ldots \hat{S}_{l_{N_m}}^{+}\prod_{l=1}^{2L}|\downarrow\rangle\right\}$ with $1 \leqslant l_{1}<l_{2}<\ldots<l_{N_m} \leqslant 2L$.
An $N_m$-magnon wave function can be written as $|\Psi\rangle=\sum_{l_{1}<l_{2}<\ldots<l_{N_m}} \psi_{l_{1}, l_{2}, \ldots, l_{N_m}}\left|l_{1} l_{2} \ldots l_{N_m}\right\rangle$.
For simplicity, we consider single-magnon excitations in a semi-infinite spin chain and ignore the energy constant $\Delta N_m-\frac{\Delta}{4}(2L-1)$ with $N_m=1$.
The symbols $v$ and $w$ are used as $v=-(J-\delta_0)$ and $w=-(J+\delta_0)$.
According to the eigenequation $(\hat{H}-E)|\Psi\rangle=0$, the relation between amplitudes at different sites is established in the following form,
\begin{equation}\label{WaveRelation}
\begin{aligned}
&w \psi_{0}+\left(-\frac{\Delta}{2}-E\right) \psi_{1}+v \psi_{2}=0 \\
&v \psi_{1}-E \psi_{2}+w \psi_{3}=0 \\
&w \psi_{2}-E \psi_{3}+v \psi_{4}=0 \\
&~~~~~~~~~~~~~~\vdots \\
&v \psi_{2 L-3}-E \psi_{2 L-2}+w \psi_{2 L-1}=0 \\
&w \psi_{2 L-2}-E \psi_{2 L-1}+v \psi_{2 L}=0.
\end{aligned}
\end{equation}
From the recurrence relation~\eqref{WaveRelation}, we directly read the transfer matrix in the site basis.
Multiplying the transfer matrices, we find that for the entire chain,
\begin{equation}
\left(\begin{array}{l}
\psi_{2 L} \\
\psi_{2 L-1}
\end{array}\right)=M^{L-1} M_{0}\left(\begin{array}{l}
\psi_{1} \\
\psi_{0}
\end{array}\right)
\end{equation}
with
\begin{equation}
M_{0}=\left(\begin{array}{cr}
\frac{\Delta / 2+E}{v} & -\frac{w}{v} \\
1 & 0
\end{array}\right), M=\left(\begin{array}{cc}
\frac{E^{2}}{v w}-\frac{w}{v} & -\frac{E}{w} \\
\frac{E}{w} & -\frac{v}{w}
\end{array}\right).
\end{equation}
The critical point $\Delta_c$ in Fig.~\ref{fig:chosendelta0}(a) of red solid line in the main text can be obtained by exploring the appearance of the left topological edge state.
The initial conditions are set as $L \rightarrow \infty, \psi_{0}=0, \psi_{1}=1$.
In the thermodynamic limit $L \rightarrow \infty$, the bulk-energies remain the same as those in the conventional SSH model,
\begin{equation}
\hat{H}_{k}=(v+w \cos k) \sigma_{x}+w \sin k \sigma_{y}
\end{equation}
with the energy
\begin{equation}
E_{k}=\pm \sqrt{v^{2}+w^{2}+2 v w \cos k}.
\end{equation}
Since the minimum energy in the upper band is $w-v$, the topological edge state appears for an energy $E_a=w-v-\delta E$ with $\delta E \rightarrow 0^{+}$.
We first deal with
\begin{equation}
\left(\begin{array}{cc}
\frac{\Delta / 2+E}{v} & -\frac{w}{v} \\
1 & 0
\end{array}\right)\left(\begin{array}{l}
\psi_{1} \\
\psi_{0}
\end{array}\right)=\left(\begin{array}{c}
\frac{\Delta / 2+E}{v} \\
1
\end{array}\right)=\Phi.
\end{equation}
By solving $M u_{\pm}=\varepsilon_{\pm} u_{\pm}$, the corresponding eigenvalues $\varepsilon_{\pm}$ and eigenstates $u_{\pm}$ are captured with $\varepsilon_{\pm}=\left(\alpha \pm \sqrt{\alpha^{2}-4}\right) / 2$ and $\varepsilon_{+} \varepsilon_{-}=1$.
Here $\alpha=\frac{E^{2}-w^{2}-v^{2}}{w v}$.
After expanding the state $\Phi$ as $\Phi=A u_{+}+B u_{-}$, the wave functions at the two ends of the spin chain are related to
\begin{equation}
\left(\begin{array}{l}
\psi_{2 L} \\
\psi_{2 L-1}
\end{array}\right)=A \varepsilon_{+}^{L-1} u_{+}+B \varepsilon_{-}^{L-1} u_{-}.
\end{equation}
Considering the energy $E_a$ of the focussed state, there exists $\varepsilon_{\pm} \approx\left(-2-r \pm \sqrt{r^{2}+4r}\right) / 2$ for $r \approx \frac{2(w-v) \delta E}{w v}$.
It has $|\varepsilon_{+}|<1$ and $|\varepsilon_{-}|>1$ and due to $r>0$,
which indicates $A=1$ and $B=0$ so that
\begin{equation}
\left(\begin{array}{l}
\psi_{2 L} \\
\psi_{2 L-1}
\end{array}\right)=\varepsilon_{+}^{L-1} u_{+}
\end{equation}
ensures the existence of the left topological edge state.
Meanwhile, it has $\Phi=u_{+}$ which allows us to obtain
\begin{equation}
\frac{\Delta / 2+E}{v}=\frac{\varepsilon_{+} w+v}{E}.
\end{equation}
Further, we obtain the following relation
\begin{equation}
\Delta=2\left[\frac{\left(-2-r+\sqrt{r^{2}+4 r}\right) w v / 2+v^{2}}{w-v-\delta E}-(w-v-\delta E)\right].
\end{equation}
When $\delta E \rightarrow 0^{+}$, the critical point yields $\Delta_c=-2w=2(J+\delta_0)$.

For the same parameters as Fig.~\ref{fig:chosendelta0}(a) of red solid line in the main text, the critical point for the appearance of the non-topological edge state is analytically obtained in the same manner with an energy $E_n=v+w-\delta E$ and $\delta E \rightarrow 0^{+}$.
It is surprising to find that the critical points of the two types of edge states are the same with $\Delta_c=2(J+\delta_0)$.
By employing the same procedure, when $(J-\delta_{0})/ (J+\delta_{0})<1$ with $\delta_0=0.5$, the critical points for the appearance of non-topological edge states and the disappearance of topological edge state are also analytically given with $\Delta_c=2(J+\delta_0)$.
Under the guidance of the critical point for the appearance of non-topological edge states, there exist transitions from trivial phase to nontrivial one, and vice versa.
The critical lines $\Delta_c=2(J+\delta_0)$ and $(J-\delta_0)/(J+\delta_0)=1$ are added to Fig.~\ref{fig:windingnumber} in the main text with red and white dashed lines, respectively.

\section{Two-magnon states} \label{appendixA}

\begin{figure}[htp]
\center
\includegraphics[width=0.45\textwidth]{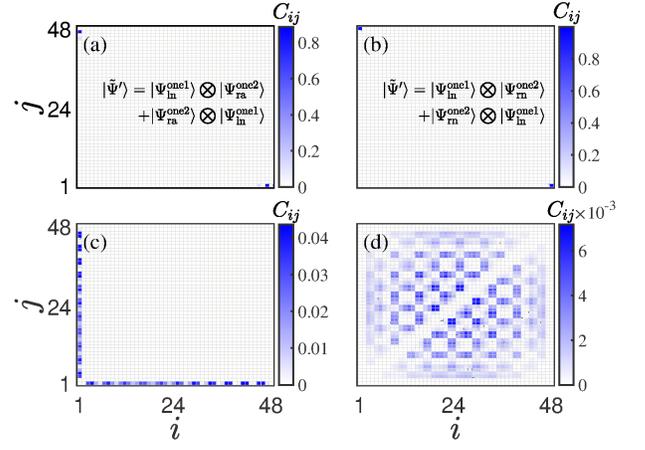}
\caption{(Color online) Two-magnon unbound states for a strong longitudinal spin-spin interaction ($\Delta=100$).
(a)-(d) are respectively correlation distributions of various focused states.
The other parameters are chosen as $J=1$, $\delta_0=-0.5$ and $L=24$.
\label{fig:TwoMagnons}}
\end{figure}

For two-magnon excitations ($N_m=2$) where two spins point up and all the other spins point down, a richer two-magnon energy spectrum can be captured under the influence of strong longitudinal spin-spin interaction, with eigenstates having the possibility of both magnon excitations locked on the same or different edges, one locked and the other free, and both free either as single magnons or as bound pairs.
Considering the two-magnon basis $\mathcal{B}^{(2)}=\left\{\left|l_{1} l_{2} \right\rangle=\hat{S}_{l_{1}}^{+} \hat{S}_{l_{2}}^{+} \prod_{l=1}^{2L}|\downarrow\rangle\right\}$ with $1\leq l_{1}<l_{2}\leq2L$, the two-magnon states can be written as $|\Psi\rangle=\sum_{l_{1}<l_{2}} \psi_{l_{1}l_{2}}\left|l_{1}l_{2}\right\rangle$.
We identify the eigenstates via the two-magnon correlation function
\begin{equation}
C_{i j}=\left\langle\Psi\left|\hat{S}_{i}^{+} \hat{S}_{j}^{+} \hat{S}_{j}^{-} \hat{S}_{i}^{-}\right| \Psi\right\rangle.
\end{equation}
$i$ and $j$ denote the lattice sites and span from $1$ to $2L$.
The two-magnon correlation functions at two specific lines $i=j\pm d$ in the $(i,j)$ plane characterize the two-magnon bound states, where $d$ relies on the specific form of the longitudinal spin-spin interaction.
Various typical unbound states are shown via the two-magnon correlations in Fig.~\ref{fig:TwoMagnons}.
$i$ and $j$ correspond to the positions of the first and second magnon excitations, and the color indicates the probability
of the two-magnon excitations occupying the $i$-th and the $j$-th sites.
Fig.~\ref{fig:TwoMagnons}(a) shows that one magnon remains localized at the left end point, while the other remains located mainly at the opposite sub-edge (the next to last site).
Fig.~\ref{fig:TwoMagnons}(b) represents an eigenstate where both magnons are located at the outermost sites, respectively.
Fig.~\ref{fig:TwoMagnons}(c) reveals that one magnon locks at the left end point while the other is extended along the chain from the third site to the next to last site.
Fig.~\ref{fig:TwoMagnons}(d) reflects that both two magnons are spread in the bulk from the second site to the next to last site.
The parameters are chosen as $J=1$, $\delta_{0}=-0.5$, $\Delta=100$ and $L=24$.

Taking the two unbound states mentioned above in Figs.~\ref{fig:TwoMagnons}(a) and (b) as examples, we understand their physical origin when extending the system from one-magnon to two-magnon excitations.
For two indistinguishable magnons, the exact two-magnon state of our open spin chain~\eqref{DimerizedSpinChain}
can be written as $|\Psi\rangle=\sum_{l_{1}<l_{2}} \psi_{l_{1}l_{2}}\left|l_{1}l_{2}\right\rangle$ where $|\psi_{l_{1}l_{2}}|^2$ describes the probability that one magnon occupies the $l_1$-th site whereas the other occupies the $l_2$-th site.
$\{\left|l_{1}l_{2}\right\rangle\}$ can viewed as two indistinguishable magnon basis with $1\leq l_1<l_2\leq2L$.
The two-magnon Hilbert space dimension is $2L(2L-1)/2$.
It allows us to expand such exact two-magnon state into the two distinguishable magnon basis as $|\tilde{\Psi}\rangle=\sum_{l_{1}l_{2}} \tilde{\psi}_{l_{1}l_{2}}\left|l_{1}l_{2}\right\rangle$ with $\tilde{\psi}_{l_{1}l_{2}}=\psi_{l_{1}l_{2}}/\sqrt{2-\delta_{l_1l_2}}$, $1\leq l_{1}\leq2L$ and $1\leq l_{2}\leq2L$.
There exists $\psi_{l_{2}l_{1}}=\psi_{l_{1}l_{2}}$.
%
%
%
Referring to the construction of higher-order topological states~\cite{2021PhysRevB104224303}, a two-magnon state $|\tilde{\Psi}'\rangle=\sum_{l_{1} l_{2}} \tilde{\psi}'_{l_{1}l_{2}}\left|l_{1} l_{2}\right\rangle$ may be constructed in terms of the single-magnon state $|\Psi^{\rm{one}}\rangle=\sum_{l_{1}}\psi^{\rm{one}}_{l_{1}}\left|l_{1}\right\rangle$ in the form
$|\tilde{\Psi}'\rangle=|\Psi^{\rm{one1}}\rangle \otimes |{\Psi}^{\rm{one2}}\rangle+|{\Psi}^{\rm{one2}}\rangle \otimes |\Psi^{\rm{one1}}\rangle$
with
\begin{equation}
\tilde{\psi}'_{l_{1}l_{2}}= \psi^{\rm{one}1}_{l_{1}}\psi^{\rm{one}2}_{l_{2}}+\psi^{\rm{one}1}_{l_{2}}\psi^{\rm{one}2}_{l_{1}}.
\end{equation}

To distinguish two types of edge states, we label left (right) non-topological edge states with $|\Psi_{\rm{ln}}\rangle$ ($|\Psi_{\rm{rn}}\rangle$)
while for left (right) topological edge states with $|\Psi_{\rm{la}}\rangle$ ($|\Psi_{\rm{ra}}\rangle$).
According to correlation properties of the two-magnon state in Fig.~\ref{fig:TwoMagnons}(a), one can observe that the first magnon is localized at the left end point and the second magnon mainly distributes at the right sub-edge site (the next to last site), or vice versa.
Combining the left non-topological edge state $\left|\Psi^{\text {one1}}_{\rm{ln}}\right\rangle$ with right topological edge state $\left|\Psi^{\text {one2}}_{\rm{ra}}\right\rangle$, we can construct such a two-magnon state with the following structure:
\begin{equation}
\tilde{\psi}^{\prime}_{l_{1} l_{2}}=\psi^{\text {one1}}_{\rm{ln},l_1} \psi^{\text {one2 }}_{\rm{ra},l_2}+\psi^{\text {one1}}_{\rm{ln},l_2} \psi^{\text {one2}}_{\rm{ra},l_1}.
\end{equation}
Our numerical calculation demonstrates the validity of the constructed two-magnon state and yields the exact two-magnon state $|\langle\tilde{\Psi}|\tilde{\Psi}^{\prime}\rangle|=1$.
This state is represented as a special type of two-magnon topological edge state where the longitudinal spin-spin interaction creates the effective potential
that traps one magnon at one end point and the other magnon forms a topological sub-edge state of the remaining sites.
Owing to the longitudinal spin-spin interaction, two magnons are oppositely distributed at one outermost site and the other sub-edge instead of bound together.
Therefore, we provide a systematic method to construct two-magnon topological edge states.
Similarly, the unbound state in Fig.~\ref{fig:TwoMagnons}(b) can also be effectively explained by the single-magnon states.
Fig.~\ref{fig:TwoMagnons}(b) corresponds to a two-magnon state with two magnons respectively locking at the two outermost sites $\left|\Psi^{\text {one1}}_{\text {ln}}\right\rangle$ and $\left|\Psi^{\text {one2}}_{\text {rn }}\right\rangle$, or vice verse.
But for the states in Figs.~\ref{fig:TwoMagnons}(c) and (d), the constructed states are invalid due to the important role of $\Delta$ in these two unbound states.
Fig.~\ref{fig:TwoMagnons}(c) shows that one magnon is edge-localized at the left end point, and the other one remains spatially extended from the third site to the next to last site due to the presence of $\Delta$, rather than fully delocalized along the chain from the second site to the next to last site.
The zero distribution on the minor diagonal lines $j=i\pm1$ in Fig.~\ref{fig:TwoMagnons}(d) implies that two magnons are free as single magnons but they cannot stay at the nearest-neighbor sites at the same time under the influence of $\Delta$.

Two-magnon excitations tend to be bound together as a whole under the influence of strong longitudinal spin-spin interaction, where two nearest-neighbor spins pointing in the
same direction with in between an arbitrary number of spins with opposite orientation are energetically bound and form a new localized effective spin.
The bound states of the magnons constitute a central part of the energy spectrum.
As a consequence, one expects a deep understanding of the interplay between the longitudinal spin-spin interaction and the transverse dimerization for strongly interacting magnons.

\begin{figure}[htp]
\center
\includegraphics[width=0.45\textwidth]{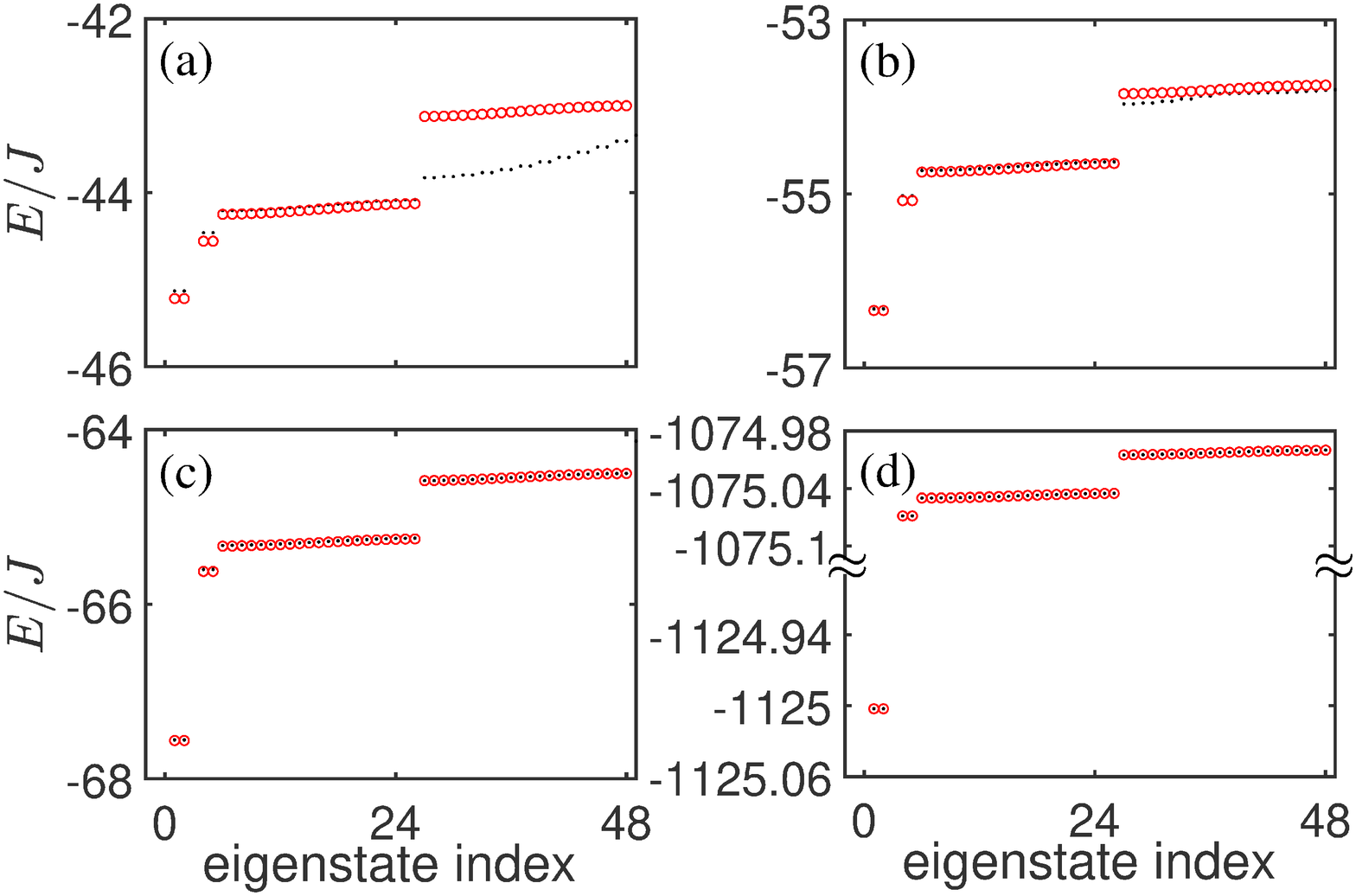}
\caption{(Color online) Two-magnon bound-state energy spectrum in ascending order for the values of the energies.
The black dots denote the energies regarding the Hamiltonian~\eqref{DimerizedSpinChain},
and the red circles are the energies obtained by the effective model $\hat{H}_{\mathrm{Eff}}^{(2)}$~\eqref{Heff2}.
The parameters are chosen as $J=1$, $\delta_{0}=-0.5$ and $L=24$, and different values of $\Delta$:
(a) $\Delta=4$, (b) $\Delta=5$, (c) $\Delta=6$, (d) $\Delta=100$.
\label{fig:ValidityTwoMagnons}}
\end{figure}

We employ the many-body perturbation theory to capture an effective model of bound magnons.
For ($|\Delta|\gg |J|, |\delta_0|$), the system~\eqref{DimerizedSpinChain} is divided into two parts $\hat{H}=\hat{H}_{0}+\hat{H}_{\text {int }}$ where the longitudinal spin-spin interaction term $\hat{H}_{\text {int }}$ plays a central role as a dominating term, while the transverse spin exchange term $\hat{H}_{0}$ is treated as a perturbation of the dominating term with
\begin{equation}
\hat{H}_{\text {int }}=-\Delta\sum_{l=1}^{2 L-1} \hat{S}_{l}^{z} \hat{S}_{l+1}^{z}
\end{equation}
and
	\begin{equation}
		\hat{H}_{0}=-\sum_{l=1}^{2 L-1}\left\{ {\left[J+(-1)^{l} \delta_0\right]\hat{S}^+_l\hat{S}^-_{l+1}+\mathrm{H.c} .}\right\}.
	\end{equation}
The dominant term $\hat{H}_{\text {int }}$ is divided into two subspaces.
The first subspace includes two classes of states: (1) the states $\{|l,l+1\rangle\}$ with $l=1,2L-1$ and $E=-\frac{3}{2}\Delta+C$,
(2) the states $\{|l,l+1\rangle\}$ with $1<l<2L-1$ and $E=-\Delta+C$.
The complementary subspace consists of (3) the states $\{|j,k\rangle\}$ with $j<k-1,j\neq1,k\neq2L$ and $E=C$,
(4) the states $\{|1,k\rangle\}$ with $2<k<2L$, $\{|j,2L\rangle\}$ with $1<j<2L-1$ and $E=-\Delta/2+C$, and (5) the states $\{|1,2L\rangle\}$ and $E=-\Delta+C$.
$C=\Delta N_m-\frac{\Delta}{4}(2L-1)$ is a constant independent on the site with $N_m=2$.
We project the system~\eqref{DimerizedSpinChain} onto the first subspace for an effective model $\hat{H}_{\mathrm{Eff}}^{(2)}=\hat{h}_{0}+\hat{h}_{1}+\hat{h}_{2}$ via a perturbation expansion up to the second order.
In the lowest order, we have
\begin{equation}
\begin{aligned}
\hat{h}_{0}=-\sum_{l=1}^{2 L-1}A_l|l, l+1\rangle\langle l, l+1|+C
\end{aligned}
\end{equation}
where $A_l=\frac{3}{2}\Delta$ for $l=1,2L-1$ and $A_l=\Delta$ for $1<l<2L-1$.
The first order $\hat{h}_1$ is equal to zero.
The second-order effective Hamiltonian reads
\begin{equation}
\begin{aligned}
\hat{h}_{2}=&-\sum_{l} B_l|l, l+1\rangle\langle l, l+1| \\
&-\sum_{l}D_l|l+1, l+2\rangle\langle l,l+1|+\text { H.c.}
\end{aligned}
\end{equation}
where $B_l=\frac{2\left[J+\delta_{0}(-1)^{l-1}\right]^{2}}{\Delta}$ for $l\neq1,2,2L-2,2L-1$,
$B_l=\frac{\left(J+\delta_{0}\right)^{2}}{\Delta}$ for $l=1,2L-1$
and $\frac{3\left(J-\delta_{0}\right)^{2}}{\Delta}$ for $l=2,2L-2$;
$D_l=\frac{3\left(J+\delta_{0}\right)\left(J-\delta_{0}\right)}{2\Delta}$ for $l=1,2L-2$ and
$D_l=\frac{\left(J+\delta_{0}\right)\left(J-\delta_{0}\right)}{\Delta}$ for $1<l<2L-2$.
After introducing the notation $\left|G_{l}\right\rangle=|l, l+1\rangle$, the effective Hamiltonian within a many-body perturbation theory up to
second order reads as
\begin{eqnarray}\label{Heff2}
\hat{H}_{\mathrm{Eff}}^{(2)}=&-&\sum_{l=1}^{2 L-1}(A_l+B_l)|G_{l}\rangle\langle G_{l}|+C \nonumber \\
&-&\sum_{l=1}^{2 L-2}D_l|G_{l+1}\rangle\langle G_{l}|+\text { H.c.}.
\end{eqnarray}
It is clearly shown that the effective potentials at the end points (the first and the last sites) and sub-edges (the second and the next to last sites) contribute to the two-magnon bound edge and sub-edge states.
The effective hopping strength $D_l$ no longer supports the SSH-type structure.
The position-dependent potential forms two continuum bands.
Therefore, two-magnon bound sub-edge states belong to non-topological sub-edge states rather than topological edge states.
Fig.~\ref{fig:ValidityTwoMagnons} shows a comparison of two-magnon bound-state energy spectra for the Hamiltonian~\eqref{DimerizedSpinChain} and the effective model $\hat{H}_{\mathrm{Eff}}^{(2)}$~\eqref{Heff2}.
When the longitudinal spin-spin interaction becomes moderately strong, the effective model $\hat{H}_{\mathrm{Eff}}^{(2)}$~\eqref{Heff2} agrees well with the two-magnon bound states in the Hamiltonian~\eqref{DimerizedSpinChain}.
For weak longitudinal spin-spin interactions, the effective model $\hat{H}_{\mathrm{Eff}}^{(2)}$~\eqref{Heff2} is not suitable for describing the magnon bound states due to the deviations.

\section{Three-magnon states} \label{appendixB}

Within three-particle Hilbert space, the energy spectrum hosts scattering states, two-body bound states, and three-body bound states.
The remarkable change in state distribution is caused by the interplay between the longitudinal spin-spin interaction and the transverse dimerization.
For the three-magnon basis $\mathcal{B}^{(3)}=\left\{\left|l_{1} l_{2} l_{3} \right\rangle=\hat{S}_{l_{1}}^{+} \hat{S}_{l_{2}}^{+} \hat{S}_{l_{3}}^{+}\prod_{l=1}^{2L}|\downarrow\rangle\right\}$ with $1\leq l_{1}<l_{2}<l_{3}\leq2L$, the three-magnon eigenstates can be expanded as $|\Psi\rangle=\sum_{l_{1}<l_{2}<l_{3}} \psi_{l_{1} l_{2} l_{3}}\left|l_{1} l_{2}l_{3}\right\rangle$.
We identify the eigenstates through the three-magnon correlation functions $C_{i j k}=\langle\Psi|\hat{S}_{i}^{+} \hat{S}_{j}^{+} \hat{S}_{k}^{+}\hat{S}_{k}^{-}\hat{S}_{j}^{-} \hat{S}_{i}^{-}| \Psi\rangle$ and spin magnetization distributions.
$i$, $j$ and $k$ respectively correspond to the positions of the first, second and third magnons, and the color represents the probability of three magnons to occupy the $i$-th, $j$-th and $k$-th sites.
The minor diagonal lines ($x=y\pm d=z\pm2d$, $x=y\pm 2d=z\pm d$, $x=y\pm d=z\mp d$) in the ($i$, $j$, $k$) space serve as a signature of the three-magnon bound states.
Specifically, $d$ depends on the longitudinal spin-spin interaction such as $d=1$ for the nearest-neighbor one in the system~\eqref{DimerizedSpinChain}.
\begin{figure}[htp]
	\center
	\includegraphics[width=0.45\textwidth]{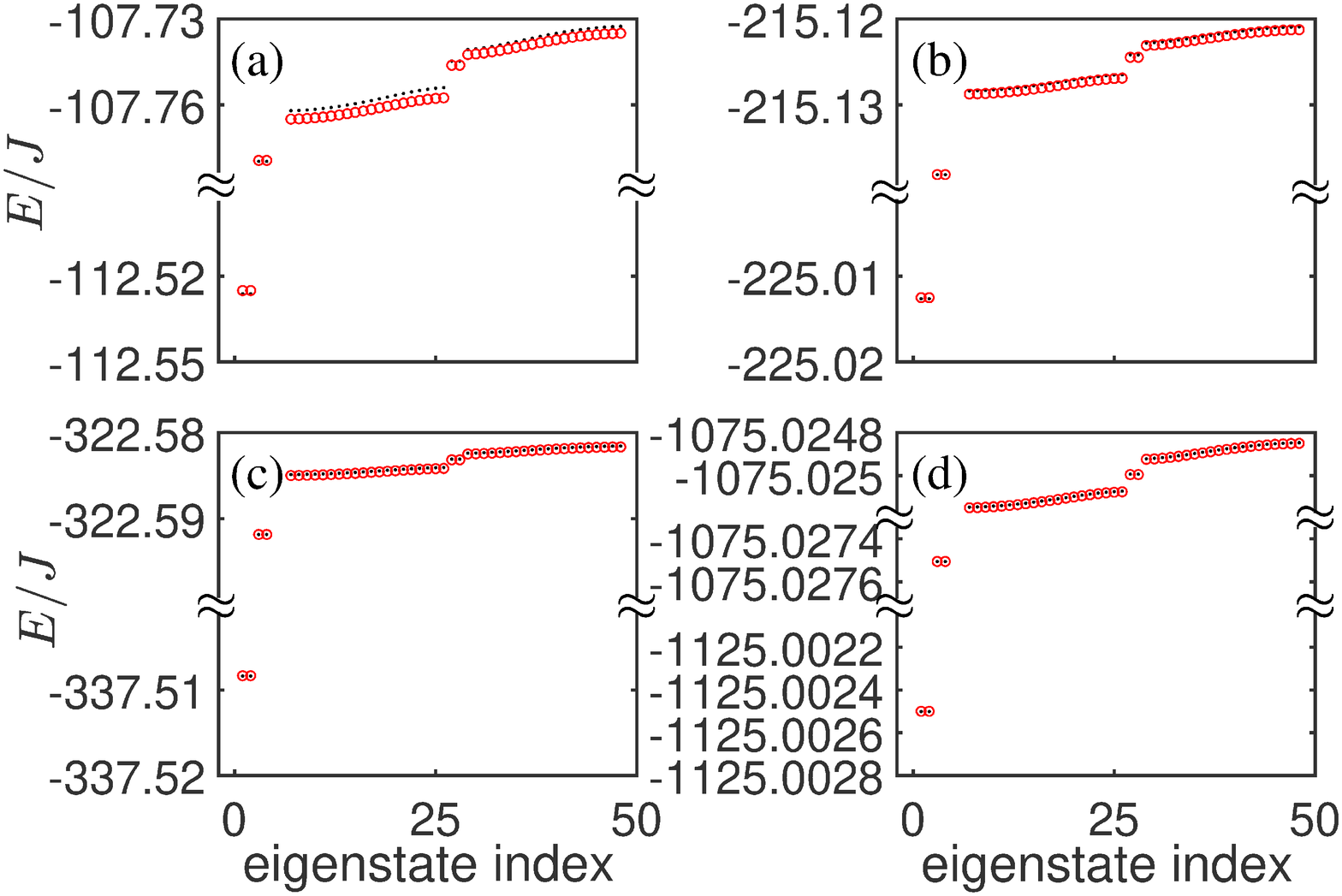}
	\caption{(Color online) Three-magnon bound-state energy spectrum in ascending order for the values of the energies.
		The black dots denote the energies regarding the Hamiltonian~\eqref{DimerizedSpinChain},
		and the red circles are the energies obtained by the effective model $\hat{H}_{\mathrm{Eff}}^{(3)}$~\eqref{Heff3}.
		The parameters are chosen as $J=1$, $\delta_{0}=0.5$ and $L=24$, and different values of $\Delta$:
		(a) $\Delta=10$, (b) $\Delta=20$, (c) $\Delta=30$, (d) $\Delta=100$.
		\label{fig:ValidityThreeMagnons}}
\end{figure}

\begin{figure*}[!htp]
	\center
	\includegraphics[width=\textwidth]{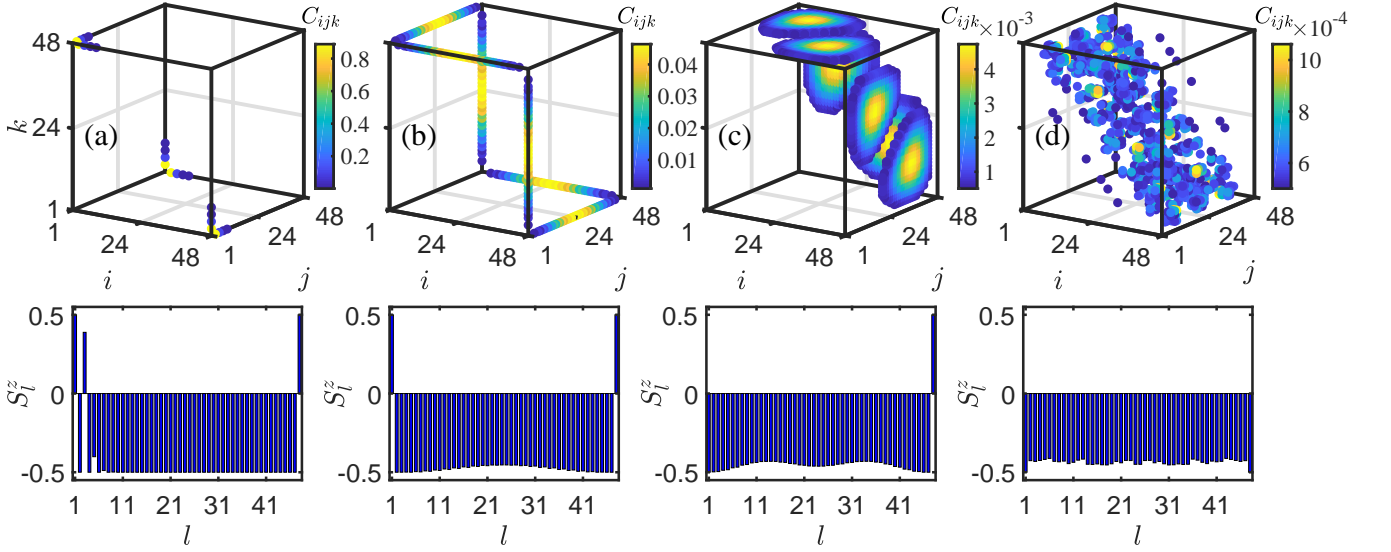}
	\caption{(Color online) Three-magnon unbound states for a strong longitudinal spin-spin interaction ($\Delta=100$).
		(a)-(d) display respectively correlation distributions of various focused states with $C_{ijk}>0.0005$.
		The other parameters are chosen as $J=1$, $\delta_0=0.5$ and $L=24$.
		The corresponding spin magnetization distributions are shown in the bottom row.
		\label{fig:ThreeSS}}
\end{figure*}

Below we first concentrate on investigating the three-magnon bound states that synchronously flip
the spins on the $l$-th, $(l+1)$-th and $(l+2)$-th sites from the ferromagnetic ground states $\prod_{l=1}^{2 L}|\downarrow\rangle$ under the strong longitudinal interaction.
In the case of the three-magnon bound states, one can refer to a detailed construction of the effective model $\hat{H}_{\mathrm{Eff}}^{(2)}$~\eqref{Heff2} to derive an effective Hamiltonian $\hat{H}_{\mathrm{Eff}}^{(3)}$~\eqref{Heff3} for the three-magnon bound states.
For a strong longitudinal spin-spin interaction, one can treat the transverse spin exchange term $\hat{H}_{0}$ as a perturbation to the longitudinal one $\hat{H}_{\text {int }}$.
The three-magnon bound-state subspace of the dominant term $\hat{H}_{\text {int }}$ is spanned by states $\left|\mathcal{G}_{l}\right\rangle=|l, l+1, l+2\rangle$ possessing $E=-5\Delta/2+\mathcal{C}$ $(l=1,2L-2)$ and $E=-2\Delta+\mathcal{C}$ $(2\leq l\leq2L-3)$.
$\mathcal{C}=\Delta N_m-\frac{\Delta}{4}(2L-1)$ is a constant independent of the lattice site with $N_m=3$.
We now project the three-magnon system~\eqref{DimerizedSpinChain} into such a bound-state subspace by implementing the perturbation analysis.
After a detailed calculation, the lowest order reads as
\begin{equation}
\hat{h}_{0}=-\sum_{l=1}^{2L-2} \mathcal{A}_l|\mathcal{G}_l\rangle\langle \mathcal{G}_l|+\mathcal{C}
\end{equation}
where $\mathcal{A}_l=\frac{5}{2}\Delta$ for $l=1,2L-2$ and $\mathcal{A}_l=2\Delta$ for $1<l<2L-2$.
The first order of the perturbation is equal to zero.
The second order of the perturbation analysis is given by
\begin{equation}
\begin{aligned}
\hat{h}_{2}=-\sum_{l=1}^{2 L-2} \mathcal{B}_l\left|\mathcal{G}_{l}\right\rangle\left\langle\mathcal{G}_{l}\right|
\end{aligned}
\end{equation}
where $\mathcal{B}_l=\frac{\left(J-\delta_{0}\right)^{2}}{\Delta}$ for $l=1,2L-2$, $\mathcal{B}_l=\frac{2\left(J-\delta_{0}\right)^{2}}{\Delta}+\frac{\left(J+\delta_{0}\right)^{2}}{\Delta}$ for $l=2,2L-3$ and $\mathcal{B}_l=\frac{\left(J-\delta_{0}\right)^{2}}{\Delta}+\frac{\left(J+\delta_{0}\right)^{2}}{\Delta}$ for $2<l<2L-3$.
Obviously, the second-order process contributes to the on-site potential especially for the distinguished effective on-site potential at the sub-edges.
The third-order perturbation reads
\begin{equation}
\begin{aligned}
\hat{h}_{3}=-\sum_{l=1}^{2 L-3} \mathcal{D}_l\left|\mathcal{G}_{l+1}\right\rangle\left\langle \mathcal{G}_{l}\right|+\text { H.c. }
\end{aligned}
\end{equation}
where $\mathcal{D}_l=\frac{5\left(J-\delta_{0}\right)\left(J+\delta_{0}\right)\left(J-\delta_{0}\right)}{2\Delta^2}$ for $l=1,2L-3$,
$\mathcal{D}_l=\frac{\left(J-\delta_{0}\right)\left(J+\delta_{0}\right)\left(J+\delta_{0}\right)}{\Delta^2}$
for $l=2,2L-4$
and
$\mathcal{D}_l=\frac{\left(J-\delta_{0}\right)\left(J+\delta_{0}\right)\left[J+\delta_{0}(-1)^l\right]}{\Delta^2}$ for $2<l<2L-4$.
The third-order term reorganizes the effective hopping strength but still supports the SSH-type structure.
At last, the three-magnon bound states obey an effective Hamiltonian up to the third order
\begin{eqnarray}\label{Heff3}
\hat{H}_{\mathrm{Eff}}^{(3)}=&-&\sum_{l=1}^{2 L-2}(\mathcal{A}_l+\mathcal{B}_l)\left|\mathcal{G}_{l}\right\rangle\left\langle \mathcal{G}_{l}\right|+\mathcal{C} \nonumber  \\
&-&\sum_{l=1}^{2 L-3} \mathcal{D}_l\left|\mathcal{G}_{l+1}\right\rangle\left\langle \mathcal{G}_{l}\right|+\text { H.c.}.
\end{eqnarray}
We can conclude that the three-magnon bound sub-edge states in Fig.~\ref{fig:Threemagnons}(c) in the main text are results of the effective on-site potential at sub-edge (the second site).
The effective SSH-type hopping strength is responsible for the three-magnon bound next-sub-edge (the third site) state in Fig.~\ref{fig:Threemagnons}(d) in the main text, that is belonged to a type of topological edge states of bound magnons.
The validity of the effective model $\hat{H}_{\mathrm{Eff}}^{(3)}$~\eqref{Heff3} is manifested in Fig.~\ref{fig:ValidityThreeMagnons}.

Apart from the three-magnon bound states discussed above, different types of three-magnon unbound states in the three-magnon energy spectrum are exhibited.
We utilize the correlation function to identify various typical three-magnon unbound states.
Compared to the bound states, there is no effective model for clarifying the three-magnon unbound states where three magnons cannot be treated as a whole.
Three unbound magnon states suffer from the problem of two of three magnons forming a bound state and unbound with the third magnon.
Therefore, the problem of how the three-magnon unbound states behave requires a further analysis.

The upper row of Fig.~\ref{fig:ThreeSS} displays correlation properties of various typical states for three unbound magnons.
Moreover, spin magnetization $S_{l}^{z}=\langle\Psi|\hat{S}_{l}^{z}|\Psi\rangle$ in the bottom row also provides an auxiliary perspective to better understand state distributions, which clearly reflects magnon distributions at each site.
The correlation function in Fig.~\ref{fig:ThreeSS}(a) reveals magnons distributed at three corners, indicating the two magnons at the two end points and the third one mainly at the third site, respectively.
Since there are two-magnon bound states, the third one is limited from the third site to the third to last site.
The non-trivial topological phase supports the topological edge states mainly distributed at the third site and the third to last site, respectively.
Meanwhile, the third magnon can also become a bulk state extending along a chain from the third site to the third to last site; see Fig.~\ref{fig:ThreeSS}(b).
Fig.~\ref{fig:ThreeSS}(c) corresponds to one magnon at the right end point and the remaining two freely distributed along a chain from the second site to the third to last site.
In addition, we observe that three magnons freely distribute in a chain from the second site to the next to last site in Fig.~\ref{fig:ThreeSS}(d).


\end{document}